\documentclass[article]{arxiv}

\usepackage{amsfonts, mathrsfs}
\usepackage{pstricks}
\usepackage{subcaption}
\usepackage{amsmath}
\usepackage{amsthm}
\usepackage{bm}
\usepackage[frozencache=true,cachedir=.]{minted} 
\usepackage{comment}
\usepackage{multirow}

\newcommand{\specialcell}[2][c]{%
  \begin{tabular}[#1]{@{}c@{}}#2\end{tabular}}

\author{Mario Beraha\\ Department of Mathematics, \\ Politecnico di Milano \\ Department of Computer Science, \\ Universit\`a di Bologna \And
        Bruno Guindani \\ Department of Electronics, Infor-\\mation and Bioengineering, \\ Politecnico di Milano \AND
        Matteo Gianella \\ Department of Mathematics, \\ Politecnico di Milano \And
        Alessandra Guglielmi \\ Department of Mathematics, \\ Politecnico di Milano}
\title{BayesMix: Bayesian Mixture Models in \proglang{C++}}

\Plainauthor{Mario Beraha, Bruno Guindani, Matteo Gianella, Alessandra Guglielmi} %% comma-separated
\Plaintitle{BayesMix: Bayesian Mixture Models in C++} %% without formatting
\Address{
  Mario Beraha\\
  Dipartimento di Matematicca\\
  Politecnico di Milano\\
  20132 Milano, Italia\\
  E-mail: \email{mario.beraha@polimi.it}
}
\newcommand{\iid}{\stackrel{\mbox{\scriptsize iid}}{\sim}}
\newcommand{\ind}{\stackrel{\mbox{\scriptsize ind}}{\sim}}

\newcommand{\virgolette}[1]{``#1''}
\renewcommand{\mid}{\,|\,}
\newcommand{\bayesmix}{\pkg{BayesMix}}
\newcommand{\bnpmix}{\pkg{BNPmix}}

\usepackage[normalem]{ulem}
\begin{document}

\begin{abstract}We describe 
\bayesmix{}, a \proglang{C++} library for MCMC posterior simulation for  general Bayesian mixture models.
The goal of \bayesmix{} is to provide  a self-contained ecosystem to perform inference for mixture models to computer scientists, statisticians and practitioners. The key idea of  this library is \emph{extensibility}, as we wish the  users to easily adapt our software to their specific Bayesian mixture models.  In addition to the several models and  MCMC algorithms for posterior inference included in the library, new users with little familiarity on mixture models and the related MCMC algorithms can extend our library with minimal coding effort. Our library is computationally very efficient when compared to competitor software. Examples show that the typical code runtimes are from two to 25 times faster than competitors for data dimension from one to ten. 
 Our library is publicly available on Github at \url{https://github.com/bayesmix-dev/bayesmix/}.
\end{abstract}

\section{Introduction}

Mixture models are a popular framework in Bayesian inference, being particularly useful for density estimation and cluster detection; see \cite{fruhwirth2019handbook} for a recent review. 
%\MBtext{Citing George Box, ``All models are wrong, but some are useful'', then 
Mixture models are convenient as they allow to decompose  complex data-generating processes into simpler pieces, for which inference is easier. Moreover, they are able  to  capture heterogeneity  and to group data together into homogeneous clusters.
The usefulness of mixture models, either finite or infinite, is evident from the huge literature developed around this topic, with applications in genomics \citep{elliott2019modeling}, healthcare \citep{beraha2022bayesian}, text mining \citep{blei2003latent} and image analysis \citep{lu202bayesian}, to cite a few. See also  
\cite{mitra2015nonparametric} for Bayesian nonparametric mixture models in biostatistical applications and the last five chapters in \cite{fruhwirth2019handbook} for applications of mixture models to different contexts, including industry, finance, and astronomy.

In a mixture model, each observation is assumed to be generated from one of $m$ groups or populations,  with $m$ finite or infinite,  and each group suitably modelled by a density, typically from a parametric family.
We consider {data} $y_1, \ldots, y_n \in \mathbb{Y} \subset \mathbb{R}^d$, $d \geq 1$.
To define a mixture model we take weights $\bm w = (w_1, \ldots, w_m)$ such that $w_h \geq 0$ for all $h = 1, \ldots, m$, $\sum_h w_h = 1$, component-specific parameters $\bm \tau = (\tau_1, \ldots, \tau_m) \in \Theta^m$, with $m<+\infty$ or $m=+\infty$, and a parametric kernel $f(\cdot \mid \cdot)$ such that $f(\cdot \mid \tau)$ is a density on $\mathbb{Y}$ for each  $\tau$ in $\Theta$ 
Specifically, we assume
\begin{equation}
\label{eq:mix_integ}
    y_i \mid \bm w, \bm \tau \iid p(y) : = \sum_{h=1}^m w_h f(y \mid \tau_h), \qquad i = 1, \ldots, n \ . 
\end{equation}
 In this paper we consider mixture models under the Bayesian approach, so that the model is completed with a prior for $(\bm w,\bm\tau)$ and $m$, i.e. 
\begin{equation}
\bm w,\bm\tau, m \sim \pi(\bm w,\bm\tau, m ) \ . 
    \label{eq:gen_prior}
\end{equation}

Posterior simulation for $(\bm w,\bm\tau, m)$ under model \eqref{eq:mix_integ}-\eqref{eq:gen_prior} is extremely challenging.
First of all, the posterior is multimodal due to the well-known label switching problem. 
Second, the number of parameters is typically huge and possibly infinite.
Several Markov chain Monte Carlo algorithms, specific for Bayesian mixture models, have been proposed since the early 2000s for posterior simulation, as, e.g.,  \cite{neal2000markov} and \cite{ishwaran2001gibbs}.
Nonetheless, as we discuss more in detail in Section~\ref{sec:software_review}, only a handful of packages are available to practitioners nowadays as, for instance, the recent \bnpmix{} \proglang{R} package \citep{ref:bnpmix} and the popular  \pkg{DPpackage} \citep{ref:dppackage}. 
%%; see Section~\ref{sec:software_review} for a more comprehensive review. 
This type of packages often provides either an \proglang{R} or a \proglang{Python} interface to some \proglang{C++} code, hence being usually efficient in fitting the associated model.

% However, \eqref{eq:mix_integ}-\eqref{eq:gen_prior} describe a very general form of mixture models and, therefore, it is impossible for a single package to fit all of them it is unrealistic that a single package could fit all of them.
 Given the generality of \eqref{eq:mix_integ}-\eqref{eq:gen_prior}, it is unrealistic to expect that a single package can be used to fit \emph{any} mixture model.
 In particular, the  choice of the parametric kernel $f(\cdot \mid \cdot)$ is  prescribed by the type of data (e.g. unidimensional vs multidimensional, continuous, categorical, counts) of the study. Many packages are built only for some type of data, and hence some kernels and priors, so that,
it is likely that  statisticians need to consider different models from the ones already available in potentially interesting software packages.
In addition, the \proglang{C++} core code is usually not written in order to be extended, with poor documentation, thus resulting in a code that is hard to make use for extensions.

To overcome these limitations, we describe here 
\bayesmix{}, a \proglang{C++} library for  Markov chain Monte Carlo (MCMC) simulation in  Bayesian nonparametric (BNP) mixture models.
The ultimate goal of \bayesmix{} is to provide statisticians a self-contained ecosystem to perform inference for mixture models. In particular, the driving idea behind this library is \emph{extensibility}, as we wish statisticians to easily adapt our software to their needs.
For instance, changing the parametric kernel $f$ in \eqref{eq:mix_integ} can be accomplished by defining a class specific to that kernel, which usually requires less than 30 lines of \proglang{C++} code. 
This new class can be seamlessly integrated in the \bayesmix{} library and, used in combination with prior distributions for the rest of the parameters and algorithms for posterior inference which are already present.
Similarly, defining a new prior for $\bm w$ requires only to implement a class for that prior, and so on. 
Therefore, new users with little familiarity on mixture models and the related MCMC algorithms can easily extend our library with minimal coding effort. 
 
The extensibility of \bayesmix{} does not come with a compromise on the efficiency. For instance, compared to  \bnpmix{} package,
%\citep{ref:bnpmix}, 
when running the same MCMC algorithm, 
our code runtimes are typically two times faster when $y_i$ is univariate and approximately 25 times faster when $y_i$ is four-dimensional.
Typical indicators of the efficiency of MCMC algorithms such as autocorrelation and effective sample size confirm that the performance obtained with our library is superior not only from the runtime point of view, but also in terms of the overall quality of the MCMC samples.
Moreover, we show that our implementation is able to scale to moderate and high dimensional settings and that \bnpmix{} fails to recover the underlying signal when $y_i$ is ten-dimensional, unlike our library.

As far as software is concerned, we achieve the desired customizability, modularity and extensibility through an object-oriented approach, making extensive use of static and runtime polymorphism through class templates and inheritance.
This may constitute a barrier for new users wishing to extend our library, as knowledge of those \proglang{C++} programming techniques is undoubtedly required.
In Section~\ref{sec:nerd_stuff} we give an example on how to implement a completely new mixture model in the library, which requires less than 130 lines of code. Then, new users can exploit this example and adapt it to their needs.

We point out that at this stage, \bayesmix{} is not \proglang{R} package, but a very powerful  and flexible \proglang{C++} library.
Although we provide a \proglang{Python} interface (see Section~\ref{sec:examples}), this is simply a wrapper around the \proglang{C++} executable.
A more sophisticated \proglang{Python} package is currently under development and available at \url{https://github.com/bayesmix-dev/pybmix}, but its description is beyond the scope of this paper.

The rest of this article is organized as follows. Section~\ref{sec:software_review} reviews software to fit Bayesian mixture models. Section~\ref{sec:algos} gives background on two  of the algorithms we have included in the library, to better understand the description of the different modules of the \bayesmix{} library in Section~\ref{sec:bayesmix}.  Section~\ref{sec:examples} shows how to install and use the library  by examples. Benchmark datasets are fitted to our library and the competitor \proglang{R} package \bnpmix{}   in Section~\ref{sec:performance}. Section~\ref{sec:nerd_stuff} contains material for more advanced users, i.e., we show how  new developers could extend the library.  The
article concludes with a discussion in Section~\ref{sec:discussion}.

\section{Review of available software} \label{sec:software_review}

One of the main drawbacks of Bayesian inference is that MCMC methods can be extremely demanding from the computational point of view. Moreover, the design of efficient MCMC algorithms and their practical implementation is not a trivial task, and thus might preclude the use of these methods to non-specialists. Nonetheless, Bayesian statistics has greatly increased in popularity in recent years, thanks to the growth of computational power of computers and the development of several dedicated software products.

In this section, we  review in particular  two packages for Bayesian mixture models, namely the \pkg{DPpackage} and the \bnpmix{} \proglang{R} packages.  They do not exhaust all the possibilities, but they are, among all software, the packages which implement the same models as in \bayesmix{}  via  the same algorithms.
%Of course, these are not the only tools that can be used to fit mixture models, but are the ones most similar to \bayesmix{}.
Other choices include  using probabilistic programming languages such as \proglang{JAGS} \citep{plummer2003jags} and \proglang{Stan} \citep{carpenter2017stan}, though their review is beyond the scope of this paper. We limit ourselves to note that \proglang{Stan} simulates from the posterior through Hamiltonian Monte Carlo while JAGS uses Gibbs sampling. \bayesmix{} uses part of the \proglang{Stan} \code{math} library for evaluating distributions, random sampling and automatic differentiation.
Observe that it is straightforward to compute the posterior of finite mixture models via \proglang{JAGS} or \proglang{Stan}. 
However, since those probabilistic programming languages work for a large class of Bayesian models, they can be less computationally efficient and fast than software purposely designed for Bayesian mixture models.

In addition to the \pkg{DPpackage} and \bnpmix{}, other \proglang{R} packages are available to fit mixture models. We report here \pkg{BNPdensity} \citep{ref:bnpdensity, ref:bnpdensity-reference} and \pkg{dirichletprocess} \citep{ref:dirichletprocess-module}.
The former focuses on nonparametric mixture models based on normalized completely random measures, using the Ferguson-Klass algorithm.
The latter focuses on Dirichlet process mixture models.
Both the packages are very flexible and implement several models and algorithms. However, they are written entirely in the \proglang{R} language, which comes as a serious drawback as far as performance is concerned.
We cite here also \pkg{NIMBLE} \citep{ref:nimble-package}, which is a hybrid between a probabilistic programming language and an \proglang{R} package, and allows to fit Dirichlet process mixture models.

%As far as the \proglang{Python} language is concerned, in addition to probabilistic programming languages, 
We also mention   the \proglang{Python} \pkg{bnpy} package \citep{ref:bnpy}, released in 2017.
The package exploits BNP models based on the Dirichlet process and finite variations of it, but forgoes traditional MCMC methods in favor of variational inference techniques such as stochastic and memoized variational inference.

The most complete software that fits BNP models is arguably the \proglang{R} library \pkg{DPpackage} \citep{ref:dppackage}.
Its most important design goal is the efficient implementation of some popular model-specific MCMC algorithms.
For this reason, it exploits embedded \proglang{C}, \proglang{C++}, and \proglang{Fortran} code for posterior sampling. 
\pkg{DPpackage} boasts a large number of features, including, but not limited to, density estimation through both marginal and conditional algorithms, ROC curve analysis, inference for censored data, binary regression, generalized additive models, and longitudinal and clustered data using generalized linear mixed models.
The Bayesian models in \pkg{DPpackage} are focused on the Dirichlet Process and its variations, e.g. DP mixtures with normal kernels, Linear Dependent DP (LDDP), Linear Dependent Poisson-Dirichlet (i.e., the Pitman-Yor mixture), weight-dependent DP, and P\'olya trees models.
Unfortunately, this package was orphaned in 2018 by its authors, and has been archived from the Comprehensive R Archive Network (CRAN) database of \proglang{R} packages in 2019.

\bnpmix{} is a recently published \proglang{R} package for Bayesian nonparametric multivariate inference \citep{ref:bnpmix}.
Its focus is on Pitman-Yor mixtures with Gaussian kernels, {thus} including the Dirichlet process mixture. This package performs density estimation and clustering through several  state-of-the-art MCMC methods, namely marginal sampling, slice sampling, and the recent importance conditional sampling, introduced by the same authors \citep{canale2020importance}. It also allows regression with categorical covariates, by using the partially exchangeable Griffiths-Milne dependent Dirichlet process (GM-DDP) model as defined in \cite{lijoi2014dependent}.

The goal of \bnpmix{} is to provide a readily usable set of functions for density estimation and clustering under a number of different BNP  Gaussian mixture models, while at the same time being highly customizable in the specification of prior information.
It also allows for different hyperpriors for the Gaussian mixture models of interest. The underlying structure of the package is written in C++, using \pkg{Armadillo} as the linear algebra library {of choice,} and {it} is integrated to \proglang{R} through the packages \pkg{Rcpp} and \pkg{RcppArmadillo}.
Inspecting the source code of \bnpmix{},  it is clear that the package lacks in modularity since, for every choice of $f(\cdot|\tau)$ and prior distribution $\pi(\bm w, \bm \tau)$,  an MCMC algorithm is implemented with little sharing of code. As a consequence, new users aiming at extending the library to other mixture models (for instance, to non-Gaussian kernels) face a tough challenge.
Since \bnpmix{} is a recent \proglang{R} package and it considers some of the mixtures our \bayesmix{} considers as well, we extensively compare the two libraries in Section \ref{sec:performance}.
However, the scopes and, probably, the end-users of \bnpmix{}  are different from those of our library as, in our opinion, 
\bnpmix{} is an \proglang{R} package providing a collection of a sort of black-box (i.e. not extensible) methods for density estimation and clustering. The \proglang{C++} functions are not documented, therefore making it difficult to extend the library to new models for new users. However, for statisticians or practitioners who 
 only intend to fit the models in \bnpmix{} to their data, this \proglang{R} package does a very good job.
%\MBtext{In our opinion, \bnpmix{} are different from those of our library is an \proglang{R} package for practitioners, that is  a collection of sort of black-box methods for density estimation and clustering, and it does very well its job. However, the \proglang{C++} functions are not documented, therefore making it difficult to extend the library to new models for new users.}

Key characteristics of good software for Bayesian mixture models thus include flexibility and the ability of providing efficient implementations of popular models.
Flexibility also comes from modularity and extensibility, as they allow re-usability of existing code, as well as combination and implementation of brand-new models and algorithms without re-writing the entire environment from scratch.
In programming terms, this often translates into the object-oriented paradigm.
These are exactly the features we have aimed at implementing into \bayesmix{}.

\section{Bayesian Mixture Models}
\label{sec:algos}

Throughout  this paper, we consider Bayesian mixture models as in \eqref{eq:mix_integ}-\eqref{eq:gen_prior}.
%of the kind \eqref{eq:mix_integ}, where $\bm w$, $\bm \tau$ and possibly $m$ are random quantities whose prior distributions are to be specified later.
For inferential purposes, it is often useful to introduce a set of latent variables $\bm c = (c_1, \ldots, c_n)$, $c_i \in \{1, \ldots, m\}$ and rewrite \eqref{eq:mix_integ} as:
\begin{equation}\label{eq:mix_clus}
\begin{aligned}
        y_i \mid \bm c, \bm \tau & \ind f(\cdot \mid \tau_{c_i}),\qquad i = 1, \ldots, n \\
        c_i \mid \bm w & \iid \text{Categorical}(\{1, \ldots, m\}, \bm w),
        \quad i = 1, \ldots, n
\end{aligned}
\end{equation}
The $c_i$'s are usually referred to as cluster allocation variables, and the clustering of the observations is the partition of $\{1, \ldots, n\}$ induced by the $c_i$'s into mutually disjoint sets $C_j = \{i: c_i = h\}$.
We refer to $m$ as the number of \emph{components} in the model, and to the cardinality of the set $\{C_j\}_j$ such that $C_j$ is non-empty as the number of \emph{clusters}. Note that the number of clusters might be strictly less then the number of components.

In the Bayesian framework, the likelihood is   complemented with   prior \eqref{eq:gen_prior} on parameters $\bm w, \bm \tau$ and possibly $m$. 
In particular, we distinguish three cases: ($i$) $m$ is finite and fixed, ($ii$) $m$ is finite almost surely but random and ($iii$) $m = +\infty$. Since $m$ can be \virgolette{large}, these mixtures are considered as belonging to the (Bayesian) nonparametric framework.
A popular choice for $f(\cdot \mid \tau)$ is the Gaussian density (unidimensional or multidimensional) with $\tau$ given by the mean and the variance (matrix). As an alternative, Student's $t$, skew-normal, location--scale or gamma densities (in case of positive data points)  might be considered. In general, the marginal prior for $\bm{w}$ is the  finite-dimensional Dirichlet distribution when $m < +\infty$ or the stick--breaking distribution when $m = +\infty$. Parameters $\tau_i$'s are typically assumed  independent and identically distributed (iid) from a suitable distribution.
The goal of the analysis is then estimating the posterior distribution of the parameters, i.e., the conditional law of $(\bm w, \bm \tau, m)$ given observations $\bm y$ (when $m$ is fixed we can consider the distribution of $m$ as a degenerate point-mass distribution).
Such posterior distribution is not available in closed form and Markov chain Monte Carlo algorithms are commonly employed to sample from it.

Of course, the algorithms for posterior inference will be different depending on  the value of $m$ (see above). 
Case ($i$) is the easiest, as a careful choice of the marginal priors for $\bm w$ and $\bm \tau$ leads to closed-form expression for the full conditionals, so that inference can be carried out through a simple Gibbs sampler.
In case ($iii$),  the whole set of parameters cannot be physically stored in a computer, and algorithms need to rely on  marginalization {techniques} \citep[see, e.g.][]{neal2000markov, walker2007sampling, papaspiliopoulos2008retrospective, kalli2011slice, griffin2011posterior, canale2020importance}.
Case ($ii$) requires a transdimensional MCMC sampler \citep{green1995reversible}, examples of which are the split-merge reversible jump MCMC \citep{richardson1997bayesian} and the birth-death Metropolis-Hastings \citep{stephens2000bayesian}  algorithm.
In the context of our work, we distinguish between \emph{marginal} and \emph{conditional} algorithms. The  former marginalize out the $m-k$ non-allocated components from the state space,  dealing only with the cluster allocations; examples are the celebrated algorithms by Neal \citep{neal2000markov}.
The  latter instead store the whole parameters state (or an approximation of it if $m=+\infty$); examples  include the Blocked-Gibbs sampler in \cite{ishwaran2001gibbs}, the retrospective sampler in \cite{papaspiliopoulos2008retrospective} and the slice sampler in \cite{walker2007sampling}.

In the remainder of this section, we present two {well-known} algorithms for posterior inference in  detail. This will be useful in Section~\ref{sec:bayesmix} to understand the modules of the \bayesmix{} library.
For observations $y_1, \ldots, y_n$ we assume the likelihood as in \eqref{eq:mix_integ} (or equivalently as in \eqref{eq:mix_clus}) and assume that $\bm w \sim \pi(\bm w)$ and $\tau_h  \iid G_0$, $h=1, \ldots, m$, where $G_{0}$ denotes a distribution over $\Theta\subset \mathbb{R}^p$, for some positive integer $p$. 

\subsection{A marginal algorithm: Neal's Algorithm 2}
\label{sec:neal_2}

\cite{neal2000markov} {proposes} several algorithms for posterior inference for Dirichlet process mixture models.
These algorithms have been later extended to work with more general models, such as Normalized Completely Random Measures mixture models \citep[see][]{favaro2013mcmc} and finite mixture models with a random number $m$ of components \citep[see][]{miller2018mixture}.

The state of the Markov chain consists of $\bm c = (c_1, \ldots, c_n)$ and $\bm \tau = (\tau_1, \ldots, \tau_k)$, $k$ denot{ing} the number of clusters, {$k \leq m$}.
The key mathematical object for this algorithm is the so-called Exchangeable Partition Probability Function \citep[EPPF,][]{pitman1995exchangeable}, that is the prior on the clusters {configurations} $\{C_1, \ldots, C_k\}$ induced by the prior on the weights $\bm w$, when $\bm w$ is marginalized out. Following \cite{pitman1995exchangeable}, {the} probability {of realization $C_1, \ldots, C_k$ depends only on their sizes, i.e., $\Phi(n_1, \ldots, n_k)$},  where $n_h$ denotes the cardinality of $C_h$.

Neal's algorithm 2 can be summarized as follows:
\begin{enumerate}
    \item Sample each cluster allocation variable $c_i$  independently from
    \begin{equation*}
        p(c_i = h \mid \cdots) \propto 
        \begin{cases}
            \Phi(n_1^{-i}, \ldots, n_h^{-i} + 1, \ldots n_k^{-i}) f(y_i \mid \tau_h) &\text{for } h=1, \ldots, k \\ 
            \Phi(n_1^{-i}, \ldots, n_h^{-i}, \ldots n_k^{-i}, 1) m(y_i) &\text{for } h=k+1
        \end{cases}
    \end{equation*}
    where $n_h^{-i}$ denotes the cardinality of the $h$-th cluster when observation $i$ is removed from the state and $m(y_i) = \int_\Theta f(y_i \mid \theta) G_0(\mathrm{d}\theta)$.
    \item Sample the cluster-specific values  independently from $p(\tau_h \mid \cdots) \propto \prod_{i: c_i=h} f(y_i \mid \tau_h) g_0(\tau_h)$.
\end{enumerate}

Observe that in Step 1., since the $m-k$ non-allocated components and the weights $\bm w$ are integrated out when updating each cluster label $c_i$, the algorithm either assigns the $i$-th observation to one of the already existing clusters, or to a {new one}.

\bayesmix{}  allows only for the so-called Gibbs type priors \citep{deblasi2013gibbs}, for which {the probability of a new cluster is}
\begin{equation}\label{eq:eppf_gibbs}
    \Phi(n_1, \ldots, n_h, \ldots n_k, 1) = f_1(k, n, \theta) \quad\text{and}\quad
    \Phi(n_1, \ldots, n_h + 1, \ldots n_k) = f_2(n_h, n, \theta),
\end{equation}
 where $\theta$ is a (possibly multidimensional) parameter governing the EPPF, $n$ is the total number of observations, {and} $k$ is the number of clusters.
The expression of $f_1$ and $f_2$ is specific of each EPPF.

\subsection[A conditional algorithm: the Blocked Gibbs sampler by Ishwaran and James (2001)]{A conditional algorithm: the Blocked Gibbs sampler by  \cite{ishwaran2001gibbs}}
\label{sec:blocked_gibbs}

In Neal's Algorithm 2 described in Section~\ref{sec:neal_2} we  can assume $m$ to be either finite or infinite, random or fixed, as long as the EPPF is available. 
For the blocked Gibbs sampler,  instead, we need to assume
a finite and fixed $m$.
%%$m < +\infty$ and $m$ fixed.

The state of the algorithm consists of $\bm c, \bm w, \bm \tau$.  The algorithm can be summarized as follows:
\begin{enumerate}
    \item sample the cluster allocations from the discrete distribution over $\{1, \ldots, m\}$ such that $p(c_i = h \mid \cdots) \propto w_h f(y_i \mid \tau_h)$  for any $i$ (independently).
    \item Sample the weights from $p(\bm w \mid \cdots) \propto \pi(\bm w) \prod_{i=1}^n w_{c_i}$.
    \item Sample the cluster-specific parameters independently from \[
    p(\tau_h \mid \cdots) \propto G_0(\tau_h) \prod_{i: c_i = h} f(y_i \mid \tau_h), \qquad \text{for any $h$.}
    \]
\end{enumerate}

\section[The BayesMix paradigm: extensibility through modularity]{The \bayesmix{} paradigm: extensibility through modularity}
\label{sec:bayesmix}

In this section, we give a general overview of the main building blocks in \bayesmix{}.
This is enough for users to understand what is happening behind the curtains.
A more detailed explanation of the software, including the class hierarchy and the application programming interfaces (API) for each class can be found in Section~\ref{sec:nerd_stuff}, where we also give a practical example on how to extend the existing code to a new model.
The complete documentation of all the functions and classes in our library can be found at \url{https://bayesmix.readthedocs.io}.

Let us examine the algorithms in Sections~\ref{sec:neal_2} and \ref{sec:blocked_gibbs}.
{Step 3 in the Blocked Gibbs sampler (Section~\ref{sec:blocked_gibbs}) and step 2 in Neal's algorithm 2 (Section~\ref{sec:neal_2})} are identical.
This step depends only on: ($i$) {the} prior $G_0$, ($ii$) the likelihood $f(\cdot \mid \cdot)$, and ($iii$) the observations $\{y_i : c_i = h\}$.  In the rest of the paper, by \textit{likelihood} $f(\cdot \mid \cdot)$ we mean  the parametric component kernel in \eqref{eq:mix_integ}.

\paragraph{The \code{Hierarchy} module}

Observe that the update of $\tau_h$ is cluster-specific, and it can be performed in parallel  over different clusters. 
This suggests that one of the main building blocks of the code must be able to represent this update. We call these classes \code{Hierarchies}, since they depend both on the prior $g_0$ and the likelihood $f(\cdot \mid \cdot)$. 
In \bayesmix{}, each choice of $G_0$ is implemented in a different \code{PriorModel} object and each choice of $f(\cdot \mid \cdot)$ in a \code{Likelihood} object, so that it is straightforward to create a new \code{Hierarchy} using one of the already implemented priors or likelihoods.
The sampling from the full conditional of $\tau_h$ is performed in an \code{Updater} class. 
When the \code{Likelihood} and \code{PriorModel} are conjugate or semi-conjugate, model-specific updaters can be used to sample from the full conditional, either  by computing it in closed form or through a Gibbs sampling step.
Alternatively, we also provide two off-the-shelf \code{Updater}s that can be used with any combination of \code{Likelihood} and \code{PriorModel}, namely the \code{RandomWalkUpdater} and the \code{MalaUpdater}. 
The former samples from the full conditional of $\tau_h$ via a random-walk Metropolis Hastings, while the latter via the Metropolis-adjusted Langevin algorithm.
To improve modularity and performance, each \code{Hierarchy} stores the \virgolette{unique} value $\tau_h$ and the observations $\bm y_h := \{y_i : c_i = h\}$ or, as it is often the case, the sufficient statistics of $\bm y_h$ needed to sample from the full conditional of $\tau_h$.
The implemented hierarchies at the time of writing are reported in Table~\ref{tab:hiers}.

\paragraph{The \code{Mixing} module}

Step 2 in Section~\ref{sec:blocked_gibbs} depends only on the prior on $\bm w$ and on the cluster allocations, while Step 1 in both Sections~\ref{sec:neal_2} and \ref{sec:blocked_gibbs} requires an interaction between the weights (or the EPPF) and the hierarchies. 
Since the steps of the two algorithms are invariant to the choice of the prior for $\bm w$, we argue that this should be a further building block of the code. 
In our code, we represent a prior on $\bm w$ and the induced EPPF in a class called \code{Mixing}.

The following \code{Mixing} classes are currently available in the library:
\begin{enumerate}
    \item \code{DirichletMixing}: {it} represents the EPPF of a Dirihclet Process \citep{ferguson1973bayesian},
    \item \code{PitYorMixing}: {it} represents the EPPF of a Pitman-Yor Process \citep{pitmanyor},
    \item \code{TruncatedSBMixing}:  the prior on $\bm w$ given by a truncated stick breaking process \citep{ishwaran2001gibbs},
    \item \code{LogitSBMixing}: the \emph{dependent} prior on $\bm w(x_i)$, $x_i$ being a given covariate vector, as in \cite{rigon_lsb}.
    \item \code{MixtureFiniteMixing}: it represents the EPPF of a finite mixture with Dirichlet-distributed weights as in \cite{miller2018mixture}.
\end{enumerate}

\paragraph{The \code{Algorithm} module}

Finally, \code{Algorithm} classes are in charge of running the MCMC simulations.
An \code{Algorithm} operates on a \code{Mixing} and several \code{Hierarchies} (or clusters), calling their appropriate update methods (and passing the appropriate data as input).

Of course, not every choice of \code{Mixing} and \code{Hierarchy} can be used in combination with all the choices of \code{Algorithm}. For instance, Neal's Algorithm 2 requires that the \code{Hierarchy} is conjugate, while the blocked Gibbs sampler requires $m$ to be finite and fixed.
Moreover, the EPPF might not be available analytically for all choices of \code{Mixing}.
Nonetheless, we argue that these are consistent building blocks that allow us to exploit the structure shared by the algorithms without introducing redundant copy-pasted code.

\begin{table}
    \centering
    \begin{tabular}{c | c | c | c }
         Class Name & $f(\cdot \mid \tau)$ & $G_0(\cdot)$ & conjugate  \\\hline\hline
         \code{NNIGHierarchy} & $\mathcal{N}(\cdot \mid \mu, \sigma^2)$ & $\mathcal{N}(\mu \mid \mu_0, \sigma^2 / \lambda) IG(\sigma^2 \mid a, b)$ & true  \\
         \code{NNxIGHierarchy} & $\mathcal{N}(\cdot \mid \mu, \sigma^2)$ & $\mathcal{N}(\mu \mid \mu_0, \sigma_0^2) IG(\sigma^2 \mid a, b)$ & false  \\
         \code{LapNIGHierarchy} & $\mbox{Laplace}(\cdot \mid \mu, \lambda)$ & $\mathcal{N}(\mu \mid \mu_0, \sigma_0^2) IG(\lambda \mid a, b)$ & false  \\
         \code{NNWHierarchy} & $\mathcal{N}_d(\cdot \mid \mu, \Sigma)$ & $\mathcal{N}_d(\mu \mid \mu_0, \Sigma / \lambda) IW(\Sigma \mid \nu, \psi)$ & true \\
         \code{LinRegUniHierarchy} & $\mathcal{N}(\cdot \mid x^t \beta, \sigma^2)$ & $\mathcal{N}_p(\beta \mid \beta_0, \sigma^2 \Lambda^{-1}) IG(\sigma^2 \mid a, b)$ & true \\
         \code{FAHierarchy} & $\mathcal{N}_p(\cdot \mid \mu, \Sigma + \Lambda \Lambda^\top)$ & \specialcell{$\mathcal{N}_p(\mu \mid \mu_0, \psi I) \mathrm{DL}(\Lambda \mid a)$ \\ $ \prod_{j=1}^p IG(\sigma^2_j | a, b)$} & false 
    \end{tabular}
    \caption{The hierarchies implemented in \bayesmix{}. $IG$ stands for the Inverse-Gamma distribution while $\mathrm{DL}$ for the Dirichlet-Laplace distribution \citep{bhattacharya2015dirichlet}.}
    \label{tab:hiers}
\end{table}

\begin{table}
    \centering
    \begin{tabular}{c | c | c | c}
         Class Name & Reference & non-conjugate & marginal  \\\hline\hline
         \code{Neal2Algorithm} & \cite{neal2000markov} & false & true \\
         \code{Neal3Algorithm} & \cite{neal2000markov} & false & true \\
         \code{Neal8Algorithm} & \cite{neal2000markov} & true & true \\
         \code{BlockedGibbsAlgorithm} & \cite{ishwaran2001gibbs} & true & false \\
         \code{SplitAndMergeAlgorithm} & \cite{jain2004split} & false & true
    \end{tabular}
    \caption{The algorithms {coded} in \bayesmix{}. {F}rom left to right: name of the class, bibliographic reference, indicator for accepting non-conjugate hierarchies, if  the mixing must implement the \emph{marginal} methods (true) or the \emph{conditional} ones (false).}
    \label{tab:algos}
\end{table}

\section{Hands on examples}
\label{sec:examples}

 Here we show how to install and use the \bayesmix{} library. The section is meant for users who are not expert  \proglang{C++} programmers and only need to use what is already included in the library. See  
Section~\ref{sec:nerd_stuff} for material aimed at more advanced users.

\subsection{Installing the BayesMix library}

We provide a handy \code{cmake} installation that automatically handles all the dependencies.
After downloading the repository from Github (\url{https://github.com/bayesmix-dev/bayesmix}), it is sufficient to build the  executables using \code{cmake}.
We provide detailed instructions below.

\paragraph{Unix-like machines}
On Unix-like machines (including those featuring macOS) it is sufficient to open the terminal and navigate to the \code{bayesmix} folder. Then the following commands
\begin{minted}{shell}
mkdir build
cd build
cmake ..
make run_mcmc
make plot_mcmc
\end{minted}
create the executables \code{run_mcmc} and \code{plot_mcmc} inside the \code{build} directory.

\paragraph{Windows machines}
 At this stage of development, Windows machines are supported only via Windows Subsystem for Linux (WSL). Hence, in order to build \bayesmix{} on Windows, you simply need to follow the instructions for Unix-like machines from the Linux terminal.

\subsection{Using the BayesMix library}
\label{sec:Section5.2}

There are two ways to interact with \bayesmix{}.  \proglang{C++} users can create an executable linking against \bayesmix{} or use (a possibly customized version of) the \code{run_mcmc} executable, which receives a list of command line arguments defining the model and the path to the data, runs the MCMC algorithm and writes the chains to a file. We give an example below. Alternatively, \proglang{Python} users can interact with \bayesmix{} via the \pkg{bayesmixpy} interface. In both cases, we consider a  Dirichlet process mixture of univariate normals, i.e.
\begin{equation}\label{eq:dpm}
\begin{aligned}
        y_1, \ldots, y_n \mid \bm w, \bm \tau & \iid \sum_{h=1}^\infty w_h \mathcal{N}(\mu_h, \sigma^2_h) \\
    w_1 = \nu_1, \quad & w_j = \nu_j \prod_{\ell < j} (1 - \nu_j), \quad j > 1\\
    \nu_j & \iid \mbox{Beta}(1, \alpha)  \\
    \tau_h := (\mu_h, \sigma^2_h) & \iid \mathcal{N}(\mu_h \mid \mu_0, \sigma_h^2/\lambda) \, \mathcal{IG}(\sigma_h^2 \mid a, b)
\end{aligned}    
\end{equation}

\subsubsection{An example via the command line}

In our code, model \eqref{eq:dpm} can be declared assuming that the mixing is the \code{DirichletMixing} class and the hierarchy is the \code{NNIGHierarchy} class. We will use algorithm \code{Neal2} for posterior simulation.
We {declare} the model using three text files.
In \code{dp_param.asciipb} we {fix} the ``total mass'' parameter of the Dirichlet process  (i.e., $\alpha$ in \eqref{eq:dpm}) to be equal to $1.0$. 
\begin{minted}{shell}
fixed_value {
    totalmass: 1.0
}
\end{minted}
In \code{g0_param.asciipb} we set the parameters of the Normal-Inverse-Gamma prior {$G_0$} as $(\mu_0, \lambda, a, b) = (0.0, 0.1, 2.0, 2.0)$:
\begin{minted}{shell}
fixed_values {
    mean: 0.0
    var_scaling: 0.1
    shape: 2.0
    scale: 2.0
}
\end{minted}
Finally, in \code{algo_param.asciipb} we specify the algorithm, {the} number of iterations (and burn-in), and {the} random seed {as follows:}
\begin{minted}{shell}
algo_id: "Neal2"
rng_seed: 20201124
iterations: 1500
burnin: 500
init_num_clusters: 3
\end{minted}
To run the executable, we call the \code{build/run_mcmc} executable with the appropriate parameters:
\begin{minted}{shell}
build/run_mcmc \
  --algo-params-file algo_param.asciipb \
  --hier-type NNIG --hier-args g0_param.asciipb \
  --mix-type DP --mix-args dp_param.asciipb \
  --coll-name chains.recordio \
  --data-file data.csv \
  --grid-file grid.csv \
  --dens-file eval_dens.csv \
  --n-cl-file numclust_chain.csv \
  --clus-file clustering_chain.csv \
  --best-clus-file best_clustering.csv
\end{minted}
where the first command line arguments are used to specify the model and algorithm.
In particular, the argument \code{---coll-name} specifies which collector to use. If it is not ``\code{memory}'', then the \code{FileCollector}  (see Section \ref{sec:io}) will be used and chains stored in the corresponding file.
The remaining arguments consist of the path to the files containing the observations (\code{---data-file}), the grid where to evaluate the predictive density (\code{---grid-file}), and the files where to store the predictive (log) density (\code{---dens-file}), the MCMC chain of the number of clusters (\code{---n-cl-file}), the MCMC chain of the cluster allocation variables (\code{---clus-file}) and the best clustering obtained by minimizing the posterior expectation of Binder's loss function (\code{---best-clus-file}). 
If any of the arguments from \code{---grid-file} to \code{---best-clus-file} is empty, the computations required to get the associated quantities are  skipped.  

\begin{figure}
    \centering
    \begin{subfigure}{0.33\textwidth}
        \includegraphics[width=\textwidth]{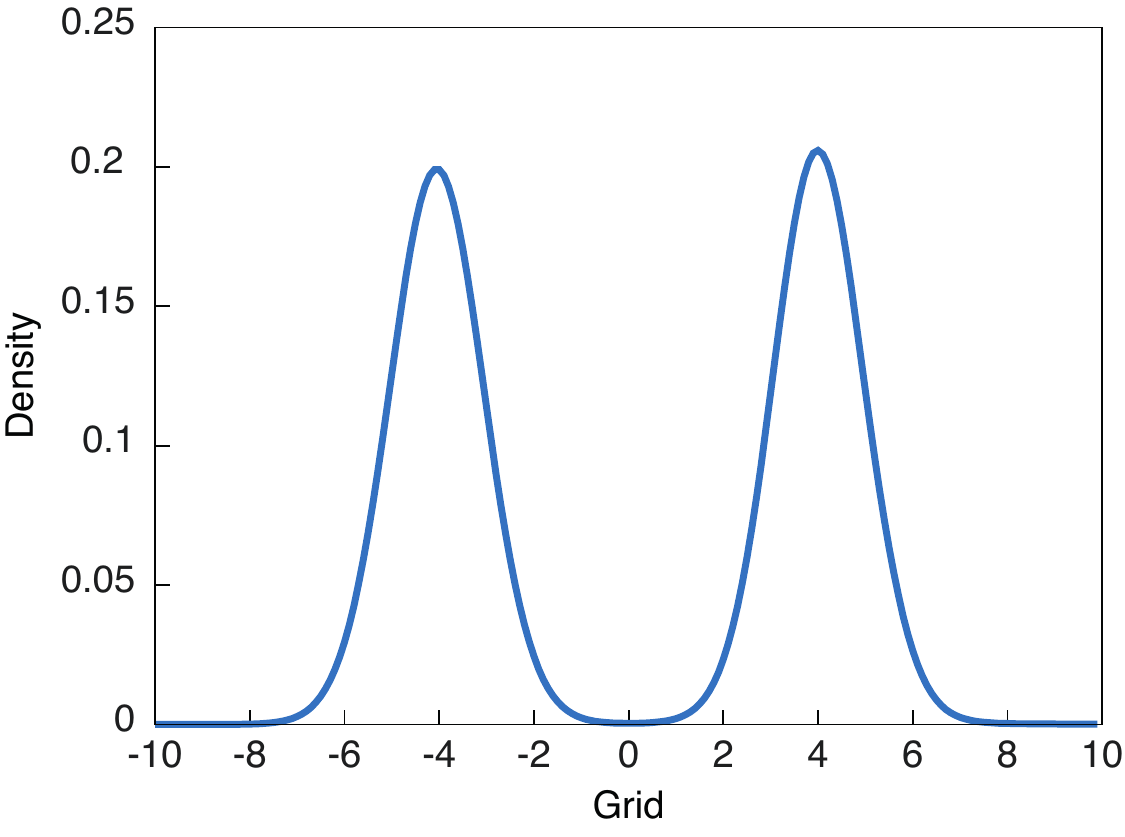}
    \end{subfigure}%
    \begin{subfigure}{0.33\textwidth}
        \includegraphics[width=\textwidth]{images/density-eps-converted-to.pdf}
    \end{subfigure}%
    \begin{subfigure}{0.33\textwidth}
        \includegraphics[width=\textwidth]{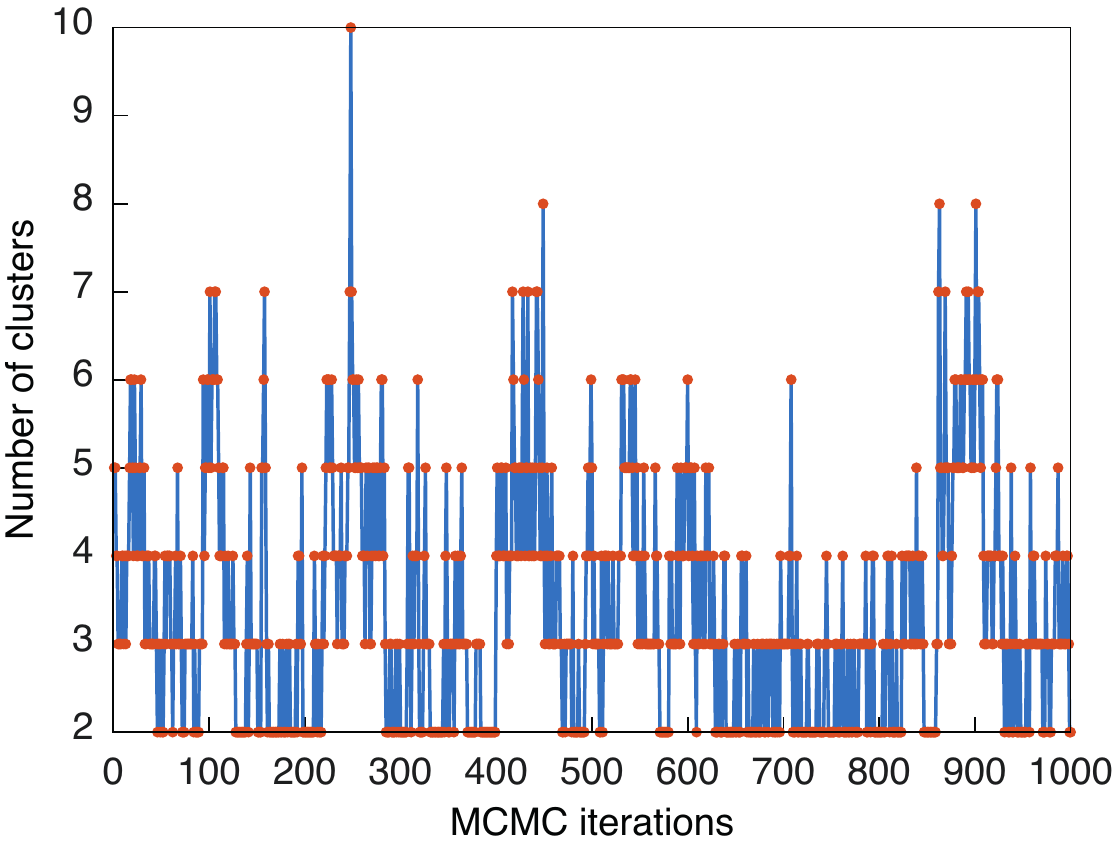}
    \end{subfigure}%
    \caption{Plots from \code{plot\_mcmc} executable: density estimate (left), histogram (center) and traceplot (right) of the number of clusters. The example refers to the \code{DirichletMixing} module  described in Section \ref{sec:Section5.2}.}
    \label{fig:commandline_plot}
\end{figure}

After the MCMC algorithm has finished to run and all the quantities of interest have been saved to \code{csv} files, it is easy to load them into another software program to summarize posterior inference through plots.
For basic uses, we provide a self-contained executable named \code{plot_mcmc} which plots and saves the posterior predictive density (Figure~\ref{fig:commandline_plot}, left panel), the posterior distribution of the number of clusters (Figure~\ref{fig:commandline_plot} (center panel)) and the traceplot of the number of clusters (Figure~\ref{fig:commandline_plot}, right panel).

\subsubsection{An example through the Python interface}

As mentioned before, we also provide (\pkg{bayesmixpy}), a \proglang{Python} interface that does not require users to use the terminal.
To install the \pkg{bayesmixpy} package, navigate to the \code{python} sub-folder and execute in the terminal ``\code{python3 -m pip install -e .}''.
Once it is installed, the package provides the \code{build_bayesmix()} and \code{run_mcmc()} functions. The former installs the executable while the latter is used to run the MCMC chains. Below, we provide a hands-on example.

First, we build \bayesmix{}:
\begin{minted}{python}
from bayesmixpy import build_bayesmix, run_mcmc
build_bayesmix(nproc=4)
>>> ...
>>>  export the environment variable BAYESMIX_EXE=<BAYESMIX_PATH>/build/run_mcmc
\end{minted}
 Observe that the last output line specifies the location of the executable and asks users to export the environmental variable \code{BAYESMIX_EXE}.
We can do it directly in Python as follows
\begin{minted}{python}
import os
os.environ["BAYESMIX_EXE"] = "<BAYESMIX_PATH>/build/run_mcmc"
\end{minted}
where \code{<BAYESMIX_PATH>/build/run_mcmc}  is the path printed by \code{build_bayesmix}.

\begin{figure}
    \centering
    \includegraphics[width=\linewidth]{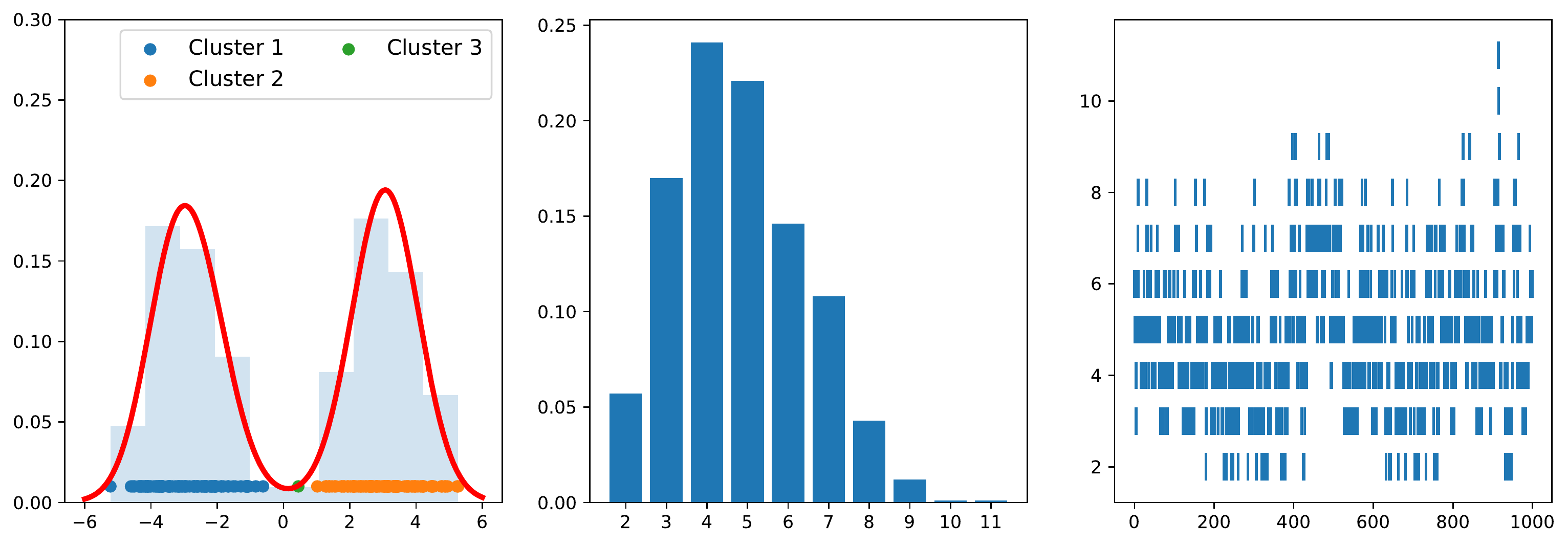}
    \caption{Output plot for the Python example: density estimate (left), histogram (center) and traceplot (right) of the number of clusters. The example refers to model \eqref{eq:dpm} described in Section \ref{sec:Section5.2}.}
    \label{fig:python_plot}
\end{figure}

We are now ready to {declare} our model. We assume a \code{DirichletMixing} as mixing and a \code{NNIGHierarchy} as hierarchy. 
The following code snippet specifies that the ``total mass'' parameter of the Dirichlet process is fixed to $1.0$, the parameters of the Normal-Inverse-Gamma prior are fixed to $(\mu_0, \lambda, a, b) = (0.0, 0.1, 2.0, 2.0)$ and we will run \code{Neal2Algorithm} for 1,500 iterations, discarding the first 500 as burn-in.
\begin{minted}{python}
dp_params = """
fixed_value {
    totalmass: 1.0
}
"""

g0_params = """
fixed_values {
    mean: 0.0
    var_scaling: 0.1
    shape: 2.0
    scale: 2.0
}
"""

algo_params = """
    algo_id: "Neal2"
    rng_seed: 20201124
    iterations: 1500
    burnin: 500
    init_num_clusters: 3
"""
\end{minted}
Finally, we  run the MCMC algorithm on some simulated data, as simply as:
\begin{minted}{python}
import numpy as np

data = np.concatenate([np.random.normal(size=100) - 3,
                       np.random.normal(size=100) + 3])
dens_grid = np.linspace(-6, 6, 1000)
log_dens, numcluschain, cluschain, bestclus = run_mcmc(
    "NNIG", "DP", data, go_params, dp_params, algo_params, 
    dens_grid=dens_grid, return_clusters=True, return_num_clusters=True,
    return_best_clus=True)
\end{minted}
which returns the log of the predictive density evaluated at \code{dens_grid} for each iteration of the MCMC sampling, the chain of the number of clusters, the chain of the cluster allocations, and the best clustering obtained by minimizing the posterior expectation of Binder's loss function.
We  summarize the inference in a plot {as follows:}
\begin{minted}{python}
import matplotlib.pyplot as plt

fig, axes = plt.subplots(nrows=1, ncols=3, figsize=(20, 5))

axes[0].hist(data, alpha=0.2, density=True)
for c in np.unique(bestclus):
    data_in_clus = data[bestclus == c]
    axes[0].scatter(data_in_clus, np.zeros_like(data_in_clus) + 0.01, 
                    label="Cluster {0}".format(int(c) + 1))
axes[0].plot(dens_grid, np.exp(np.mean(log_dens, axis=0)), color="red", lw=3)
axes[0].legend(fontsize=16, ncol=2, loc=1)
axes[0].set_ylim(0, 0.3)


x, y = np.unique(numcluschain, return_counts=True)
axes[1].bar(x, y / y.sum())
axes[1].set_xticks(x)

axes[2].vlines(np.arange(len(numcluschain)), numcluschain-0.3, numcluschain+0.3)
plt.show()
\end{minted}
The output of the above code is displayed in Figure \ref{fig:python_plot}.

\begin{figure}
    \centering
    \includegraphics[width=\linewidth]{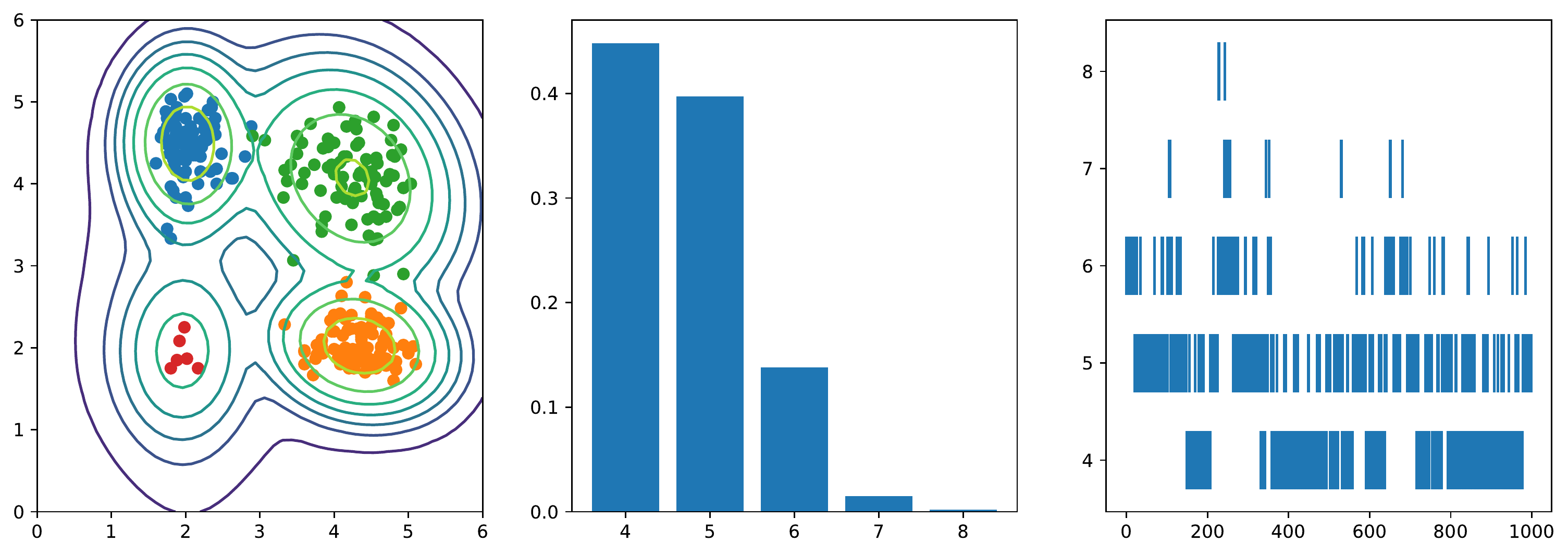}
    \caption{density estimate (left), histogram (center) and traceplot (right) of the number of clusters. The example refers to the \code{faithful} dataset in Section \ref{sec:Section5.2}.}
    \label{fig:python_plot2d}
\end{figure}

We also consider an example with bivariate datapoints, the \code{faithful} dataset, a well-known benchmark dataset for Bayesian density estimation and cluster detection.
In this case, we assume that $f(\cdot \mid \tau)$ is the bivariate Gaussian density, with parameters $\tau = (\mu, \Psi = \Sigma^{-1})$ being the mean and precision matrix, respectively.
A suitable prior for $\mu, \Psi$ is the Normal-Wishart distribution, i.e. $\mu \mid \Psi \sim \mathcal{N}_2(\mu_0, (\lambda \Psi)^{-1})$, $\Psi \sim IW(\nu_0, \Psi_0)$, with $\E(\Psi)= \Psi_0 / (\nu - 2 - 1)$.
To declare the model and run the MCMC algorithm, we can reuse most of the code of the univariate example, replacing the defintion of \code{g0_params} with:
\begin{minted}{python}
g0_params = """
fixed_values {
    mean {
        size: 2
        data: [3.484, 3.487]
    }
    var_scaling: 0.01
    deg_free: 5
    scale {
        rows: 2
        cols: 2
        data: [1.0, 0.0, 0.0, 1.0]
        rowmajor: false
    }
}
"""
\end{minted}
Posterior inference is summarized in Figure~\ref{fig:python_plot2d}.

\section{Performance benchmarking and comparisons}\label{sec:performance}

Here we compare the library \bayesmix{} and the recently published \bnpmix{} R package, which we have reviewed in Section \ref{sec:software_review}, in terms of clustering quality and computational efficiency.
All simulations were run on a Ubuntu 21.10 16 GB laptop machine.
We consider three benchmark datasets for the comparison.
The first two are the popular univariate \code{galaxy} and bivariate \code{faithful} datasets, both available in \proglang{R}.
The third example is a simulated four-dimensional dataset, which we will refer to as \code{highdim}.
It includes 10,000 points sampled from a Gaussian mixture with two equally weighted components, with mean $\mu_4 = [2,2,2,2]$ and $-\mu_4$ respectively, and both covariance matrices equal to the identity matrix.

Since \bnpmix{} focuses on Pitman-Yor processes and does not implement the Gamma prior for the total mass of the Dirichlet process, comparison is made using only Pitman-Yor mixtures with the same hyperparameter values for both libraries, including Pitman-Yor parameters and hierarchy hyperprior values.
We test \bayesmix{}  using four different marginal algorithms -- \code{Neal2}, \code{Neal3}, \code{Neal8}, and \code{SplitMerge}. The package \bnpmix{} uses its own implementation of \code{Neal2}, which is referred to as \code{mar}, and the authors' newly implemented importance conditional sampler, or \code{ics} for short.
Each algorithm has been run for 5,000 iterations, with 1,000 iterations as burn-in period.

\begin{figure}
        \centering
        \includegraphics[width=0.32\textwidth]{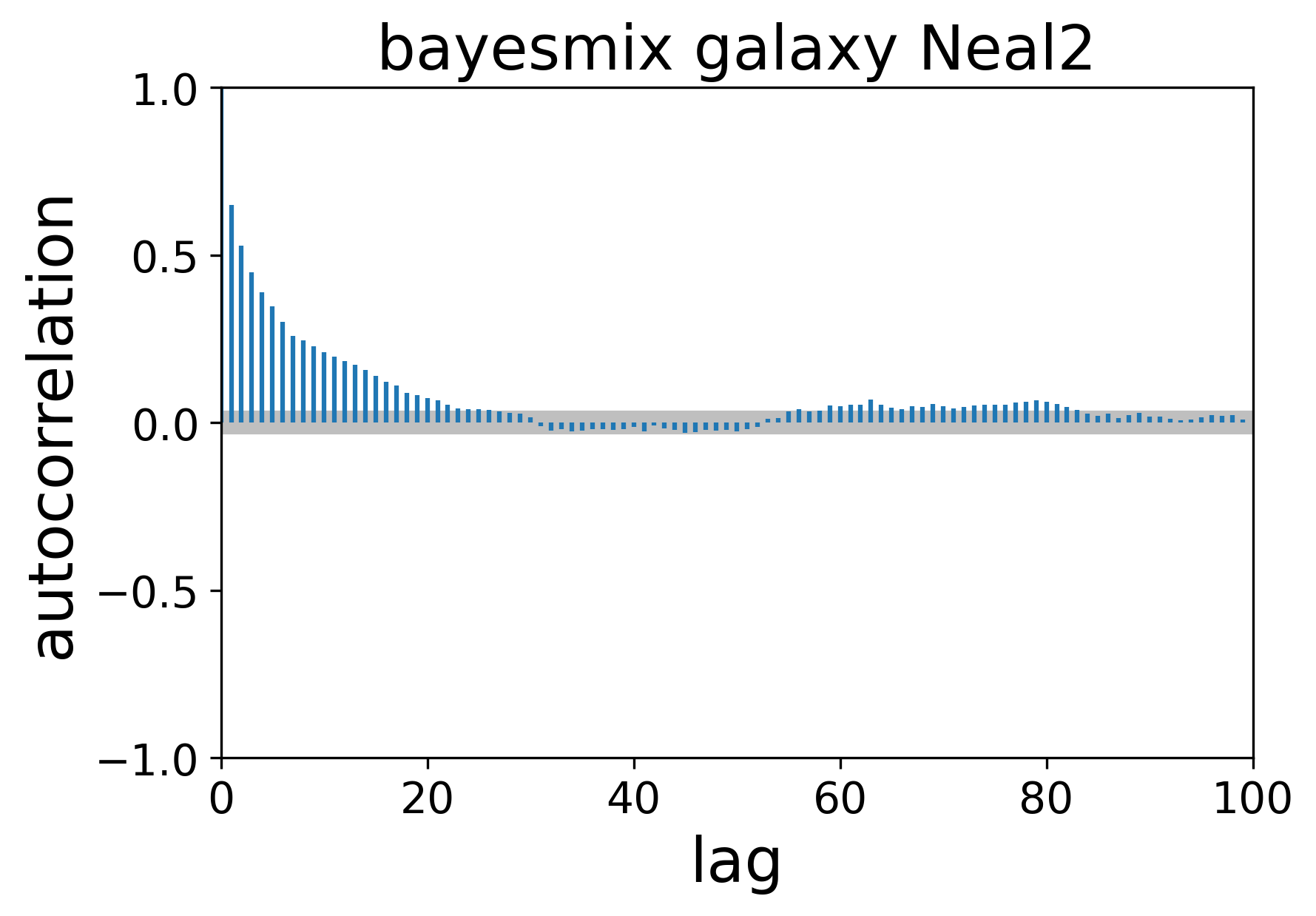}
        \includegraphics[width=0.32\textwidth]{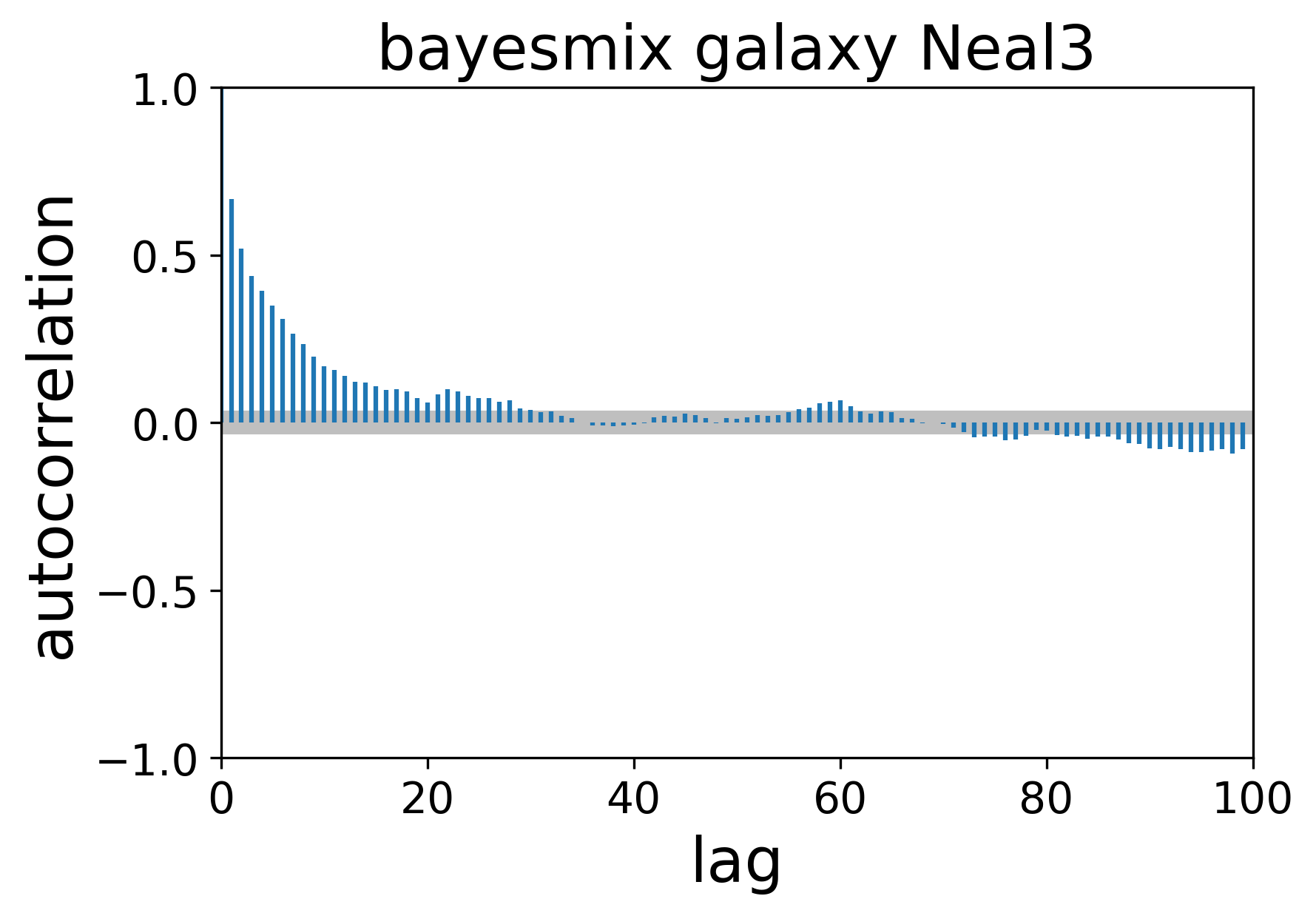}
        \includegraphics[width=0.32\textwidth]{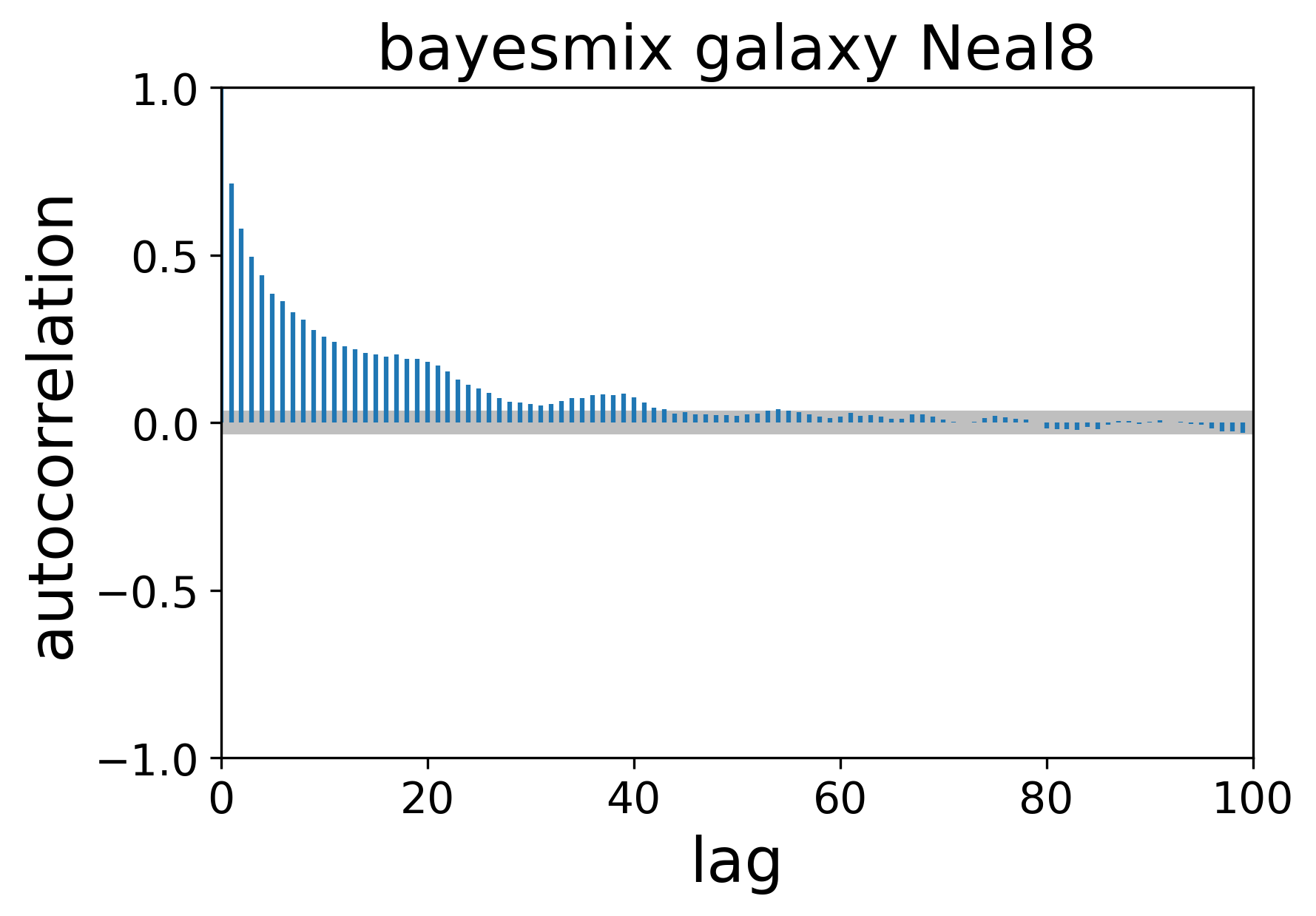}

        \includegraphics[width=0.32\textwidth]{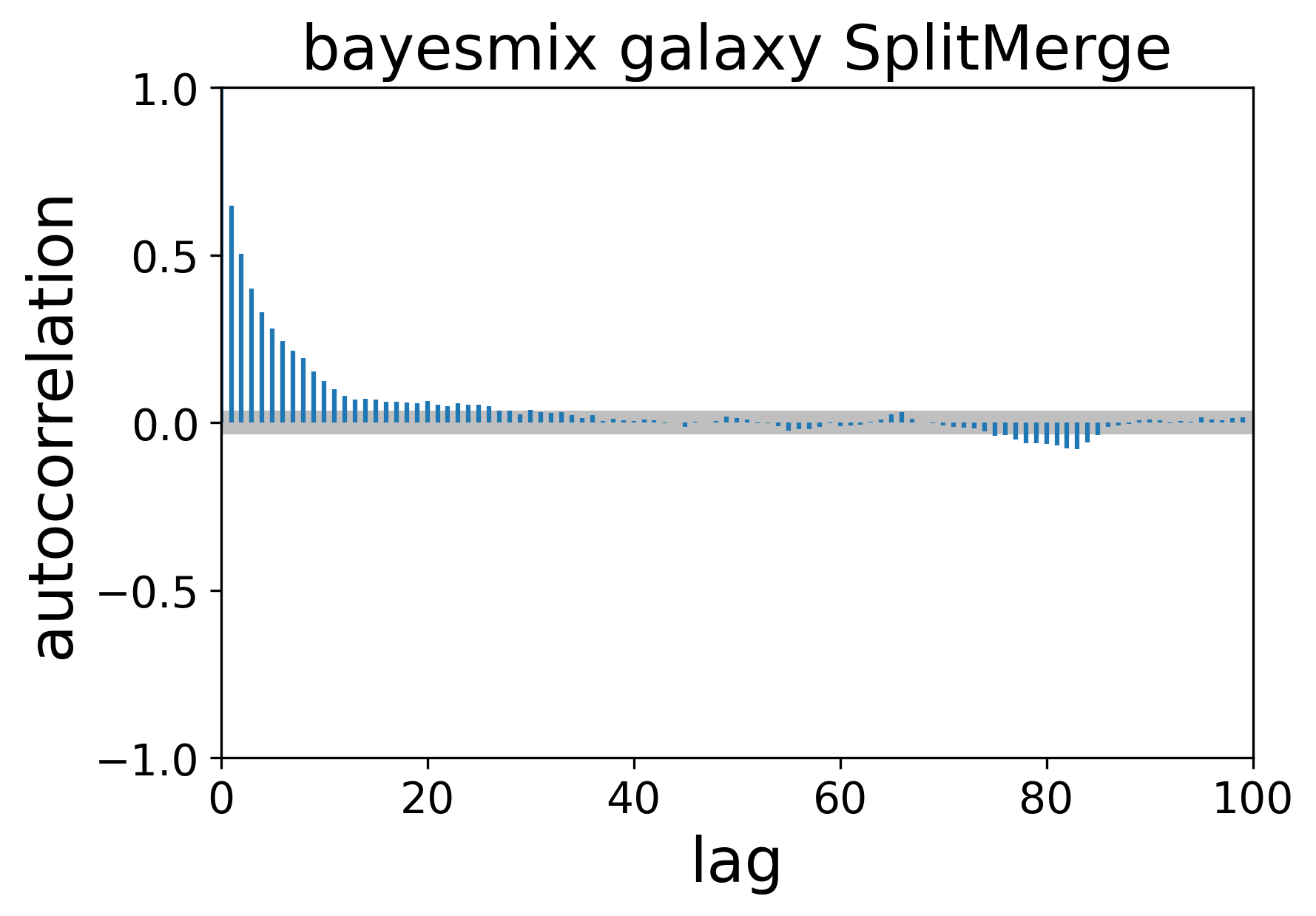}
        \includegraphics[width=0.32\textwidth]{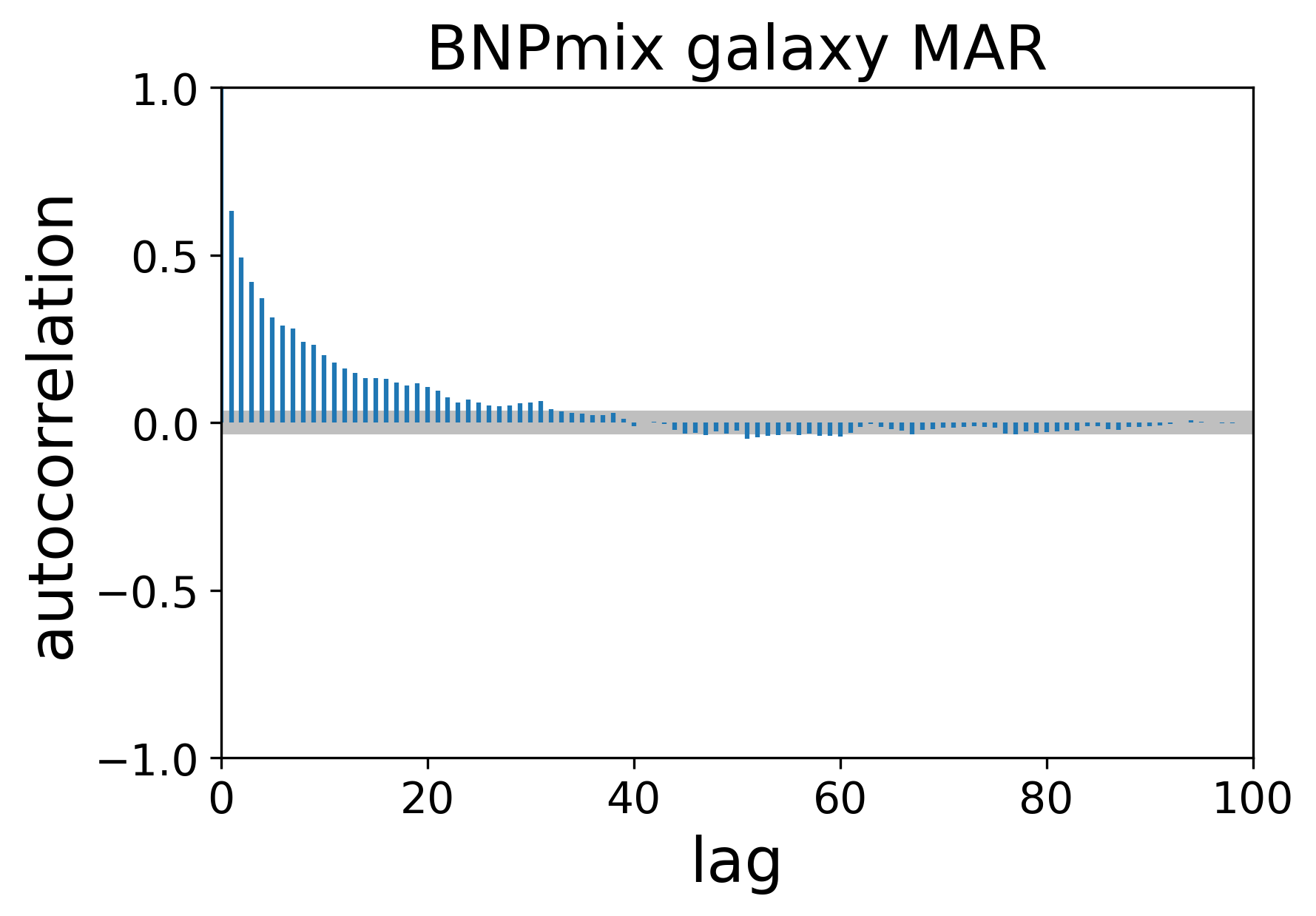}
        \includegraphics[width=0.32\textwidth]{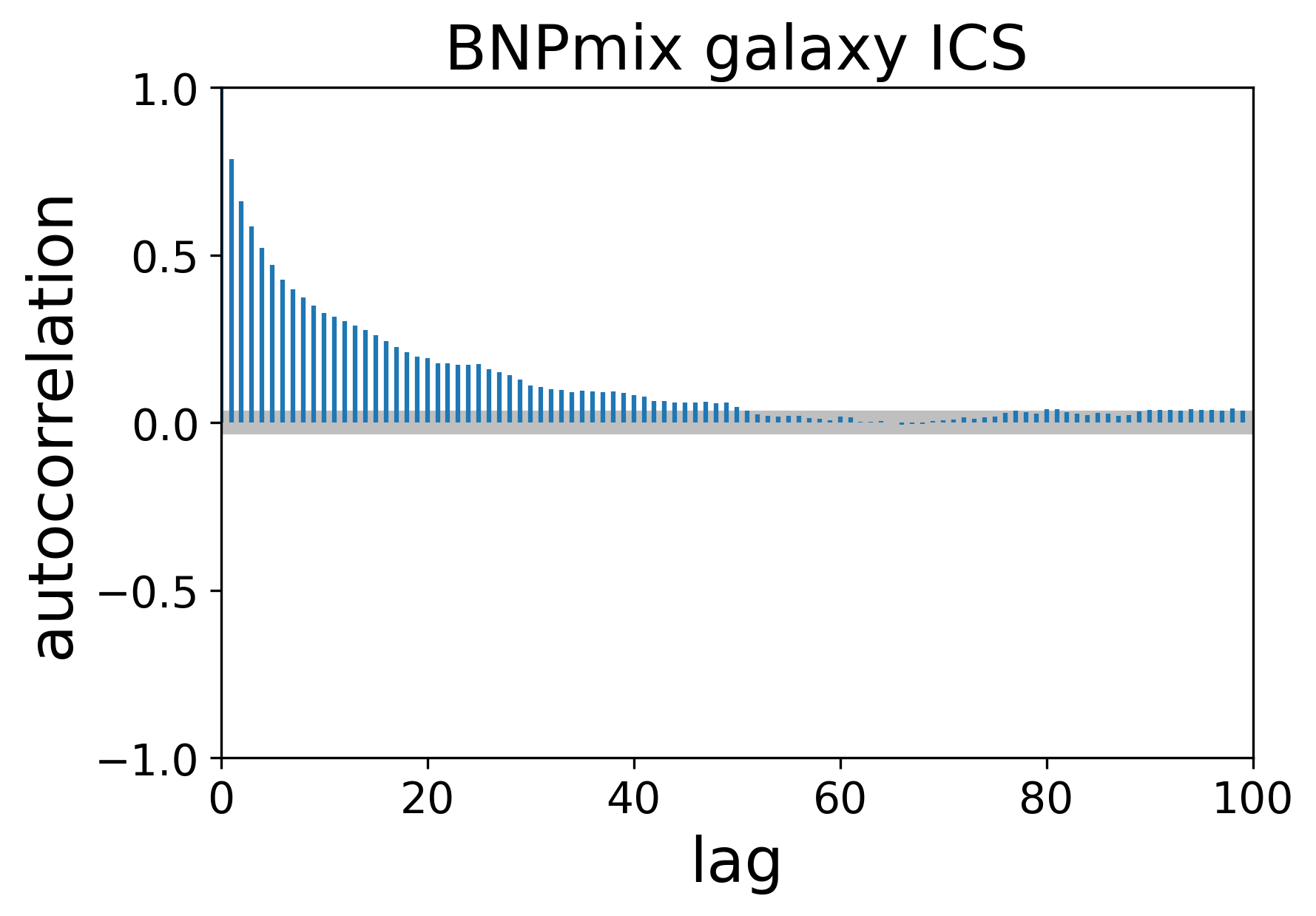}

        \includegraphics[width=0.32\textwidth]{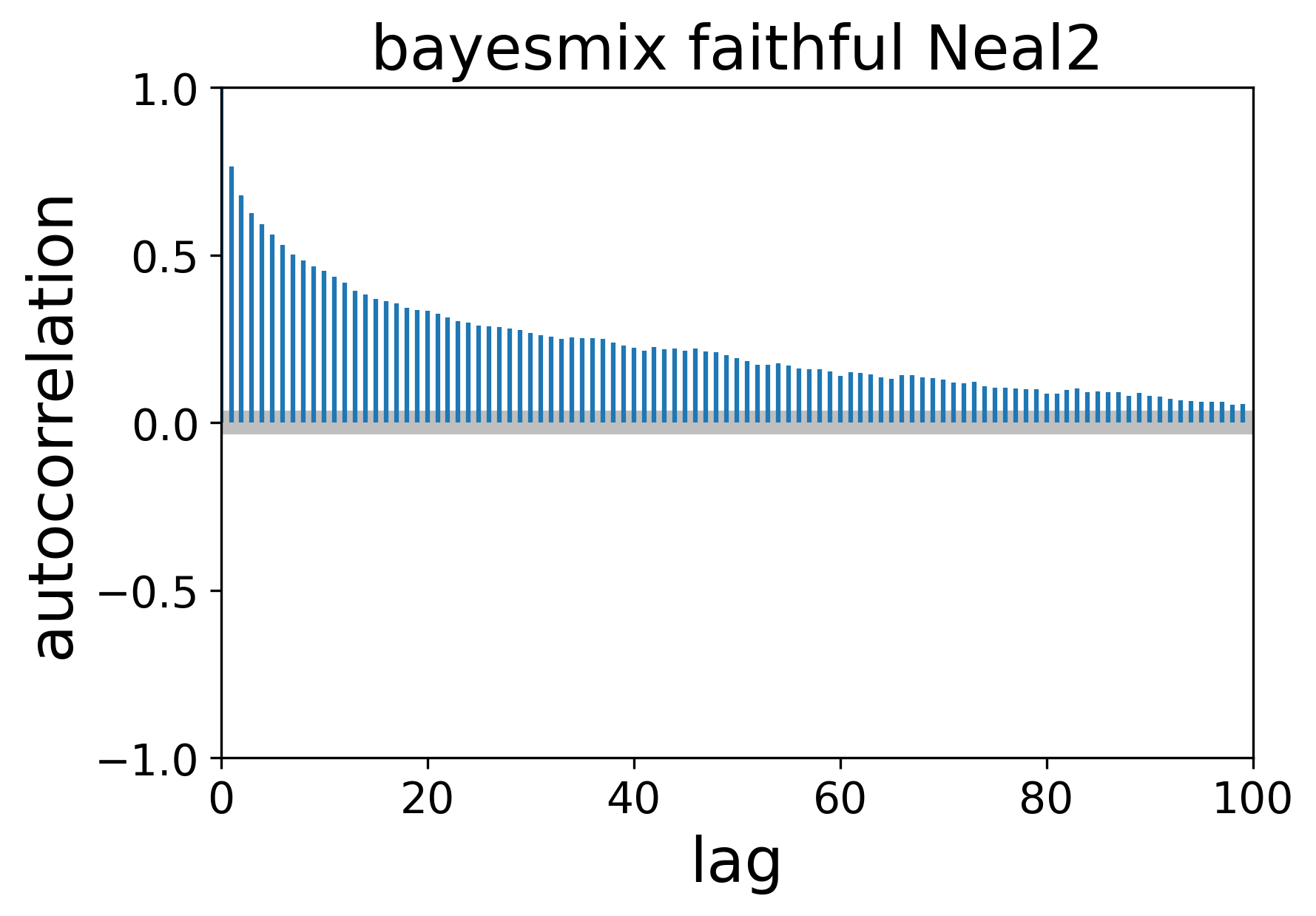}
        \includegraphics[width=0.32\textwidth]{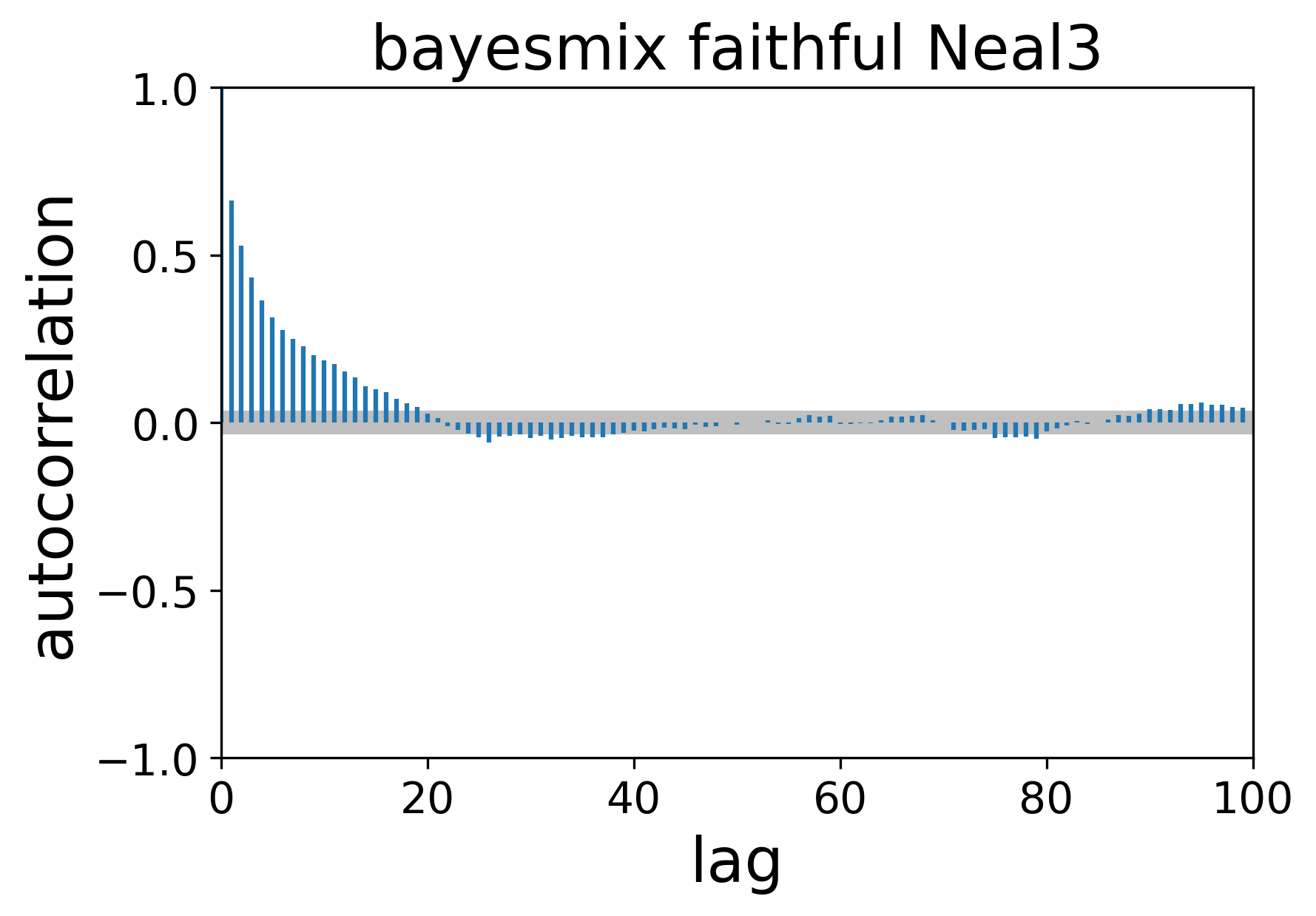}
        \includegraphics[width=0.32\textwidth]{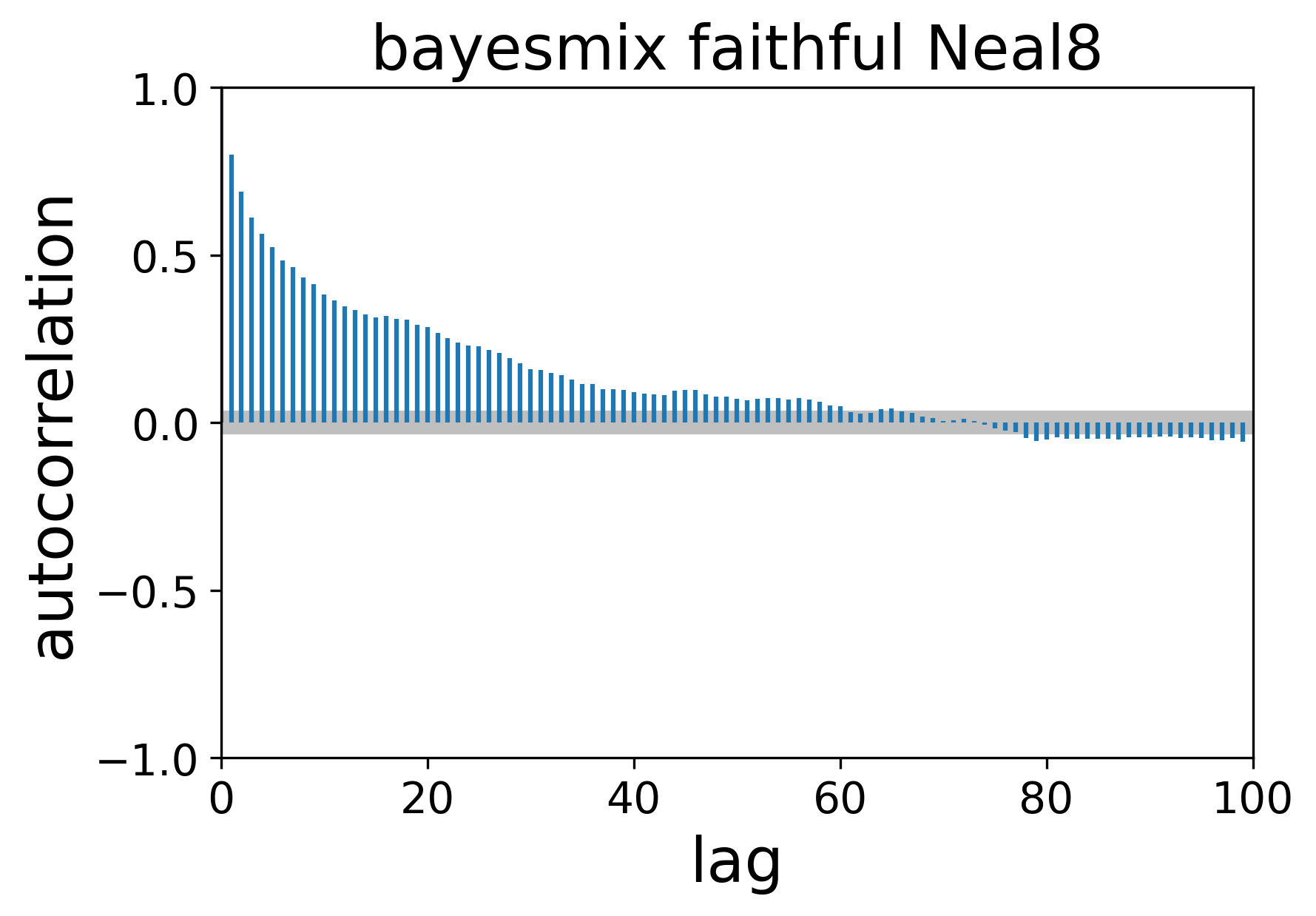}

        \includegraphics[width=0.32\textwidth]{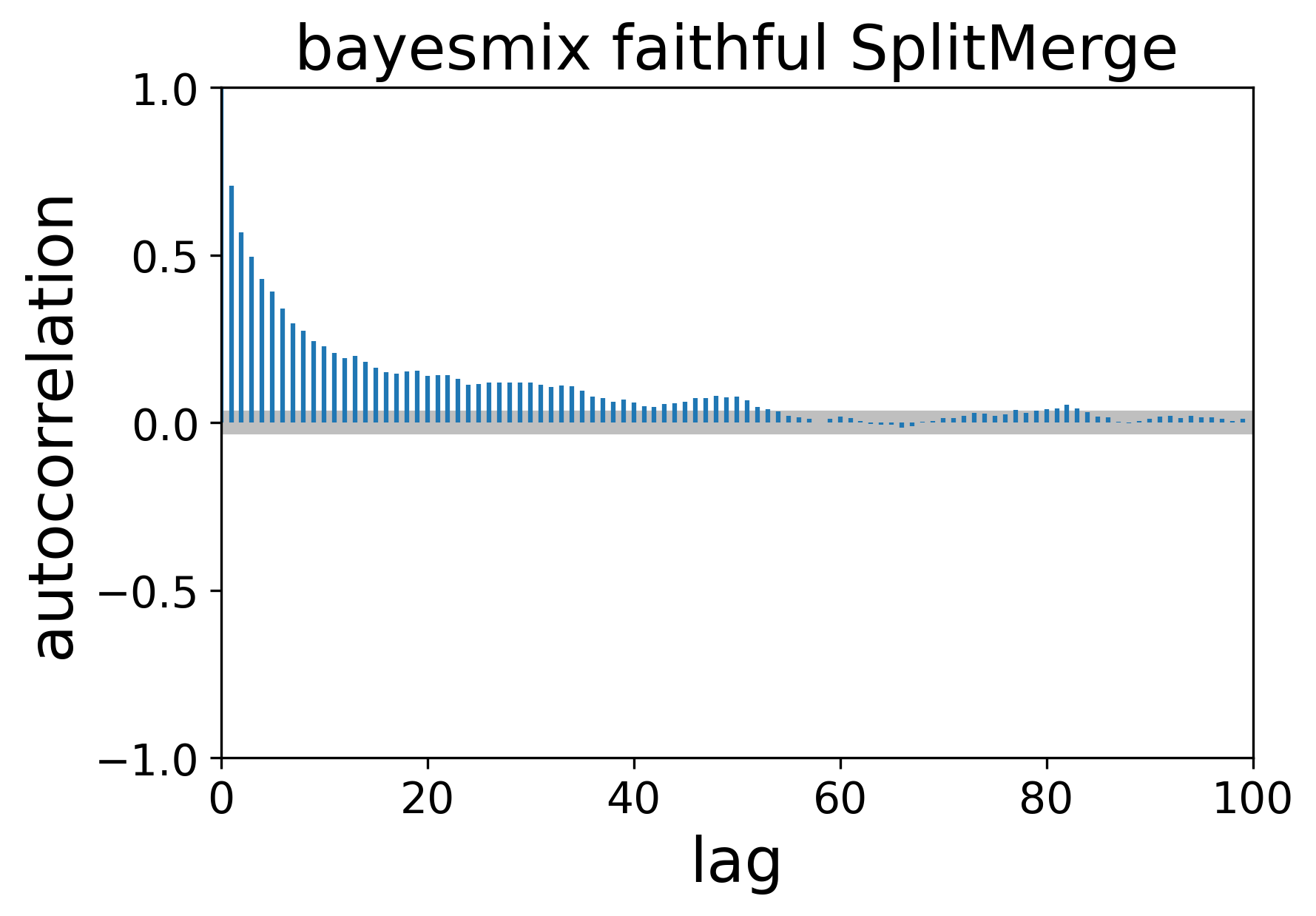}
        \includegraphics[width=0.32\textwidth]{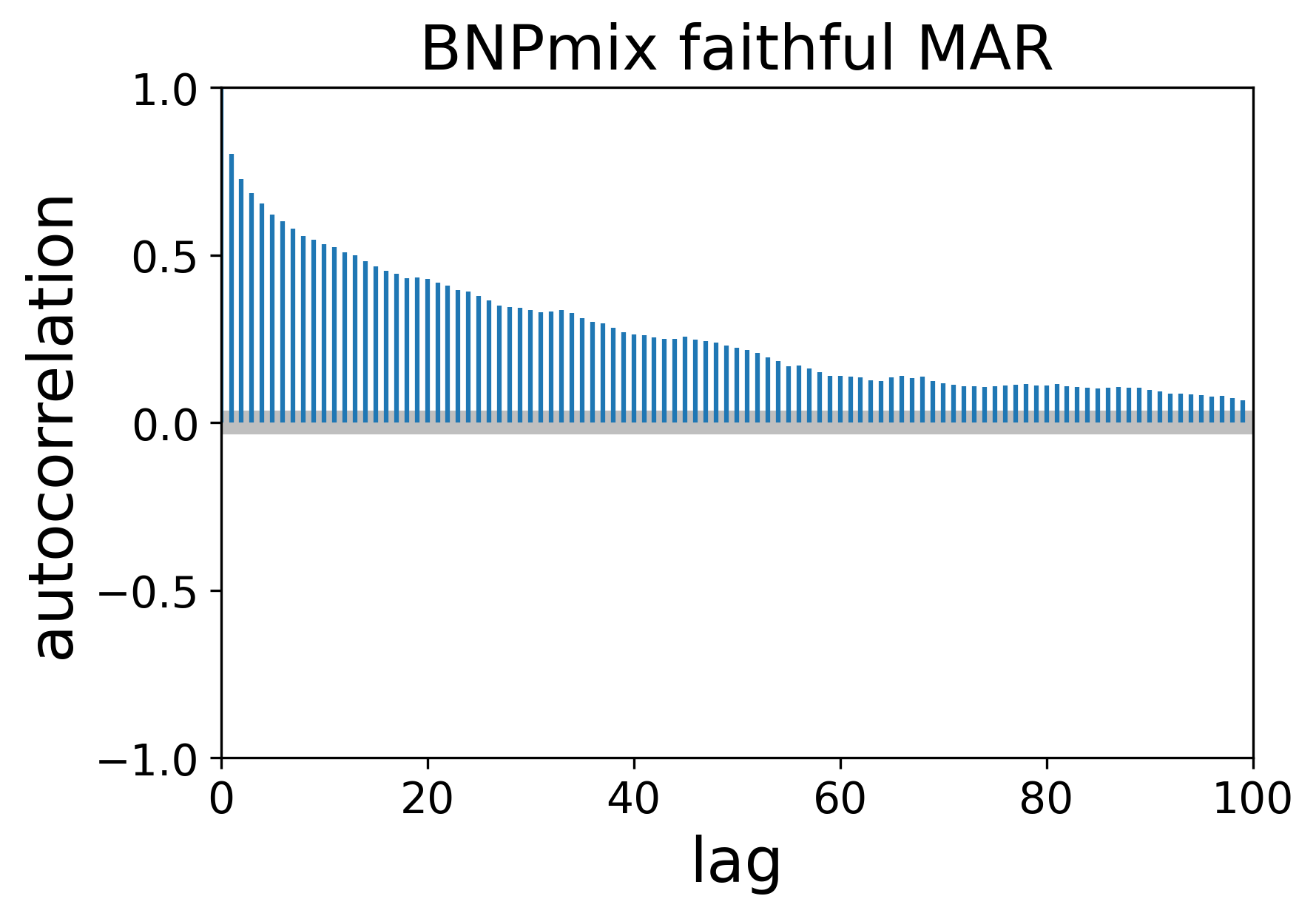}
        \includegraphics[width=0.32\textwidth]{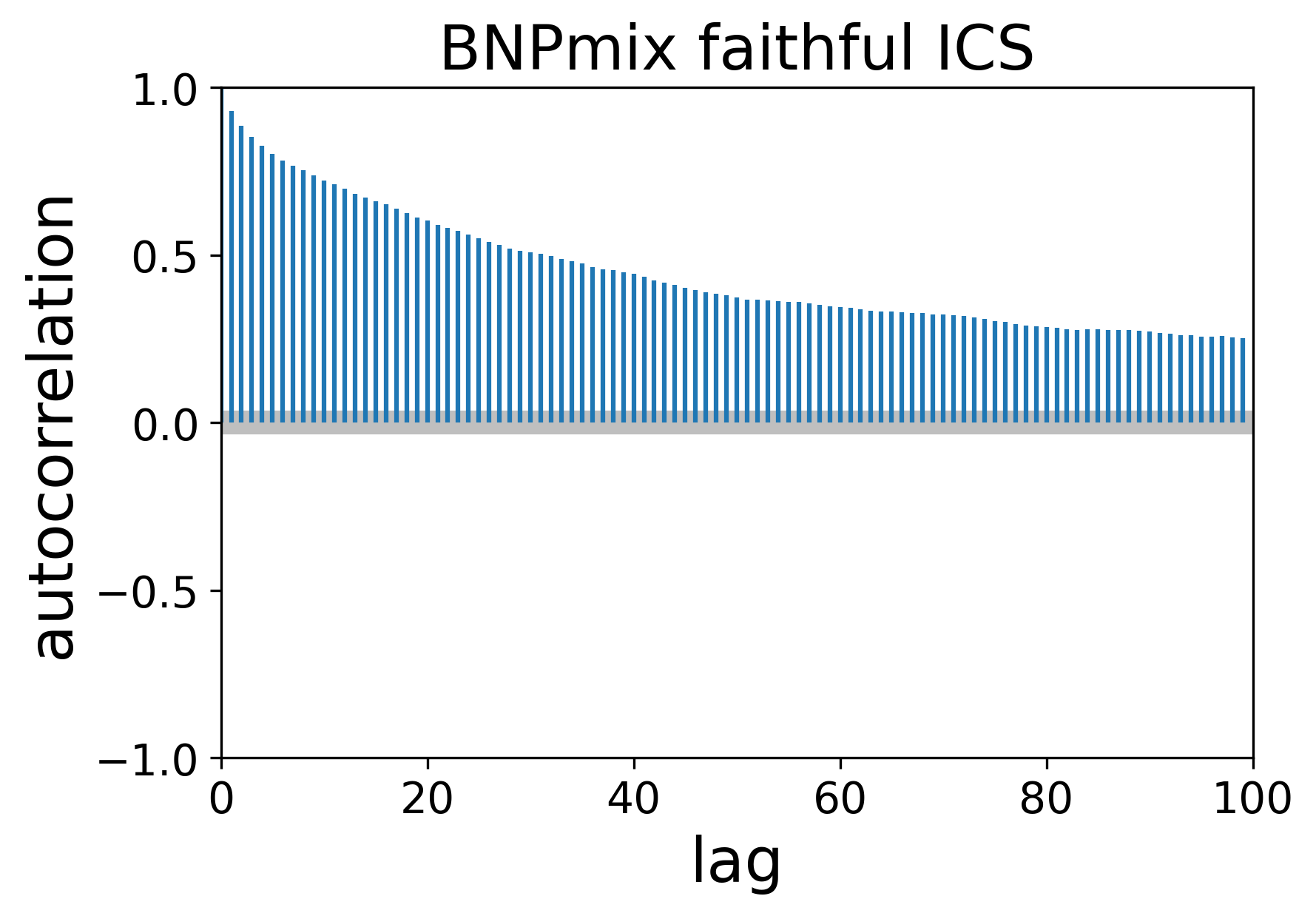}

        \includegraphics[width=0.32\textwidth]{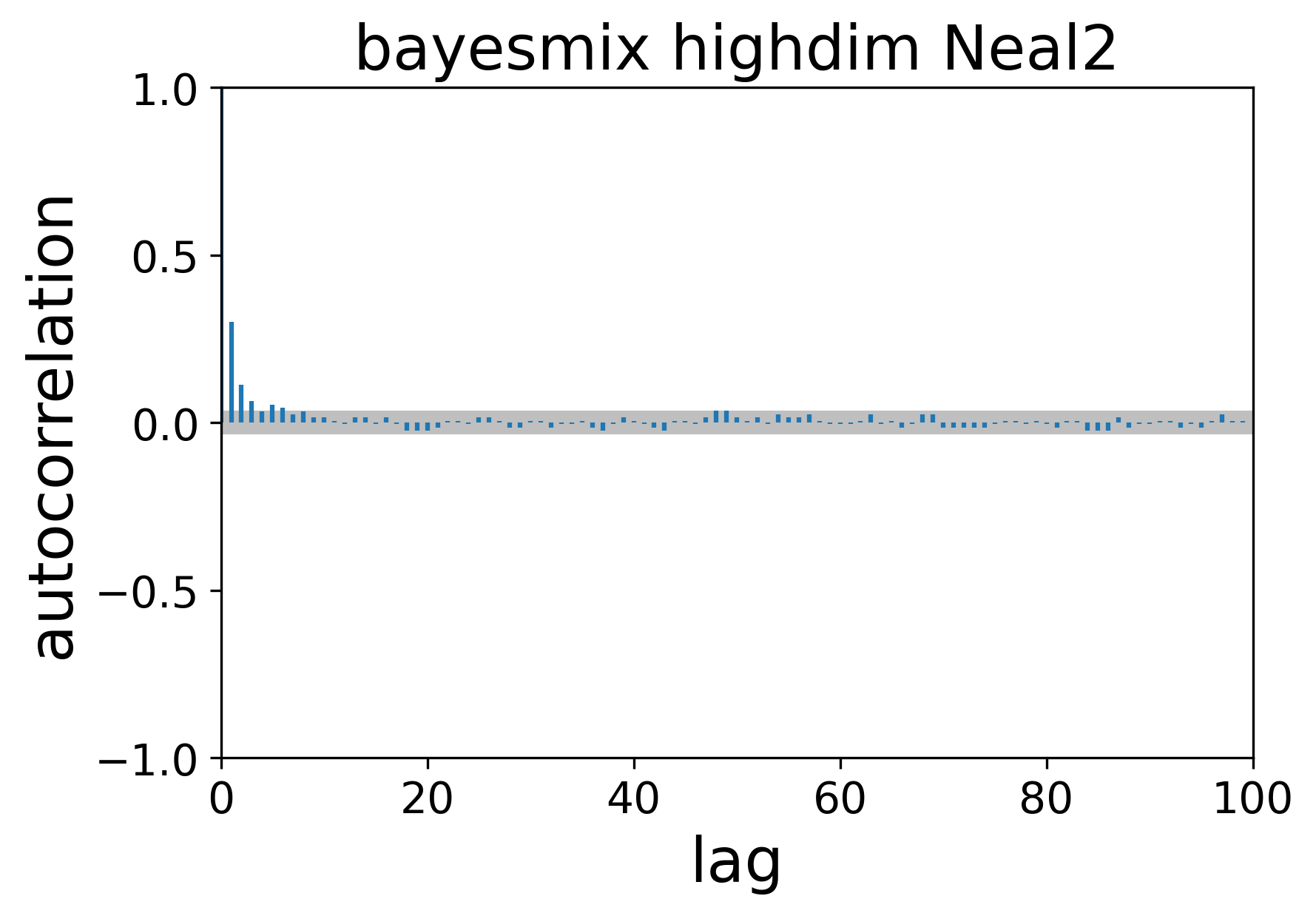}
        \includegraphics[width=0.32\textwidth]{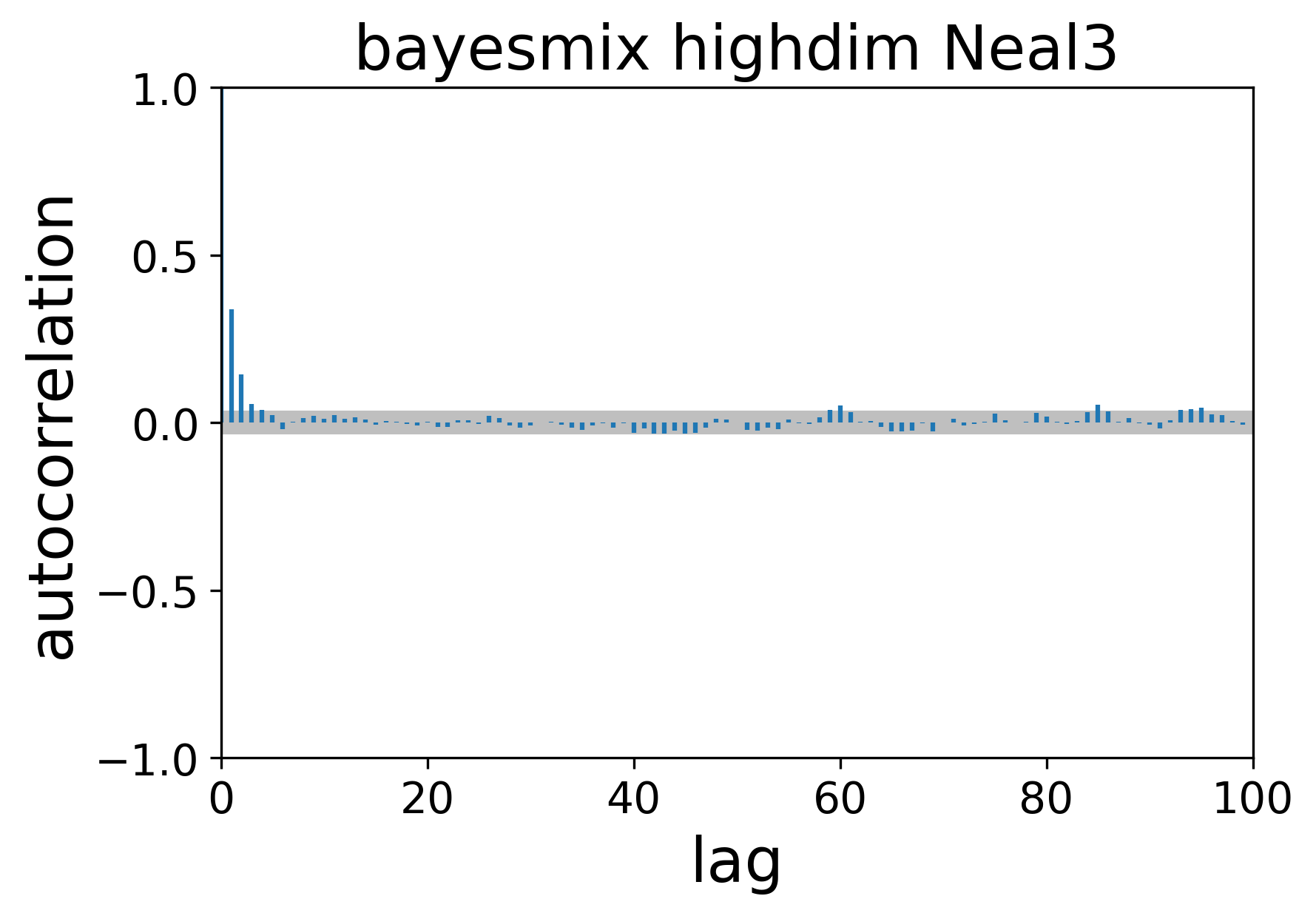}
        \includegraphics[width=0.32\textwidth]{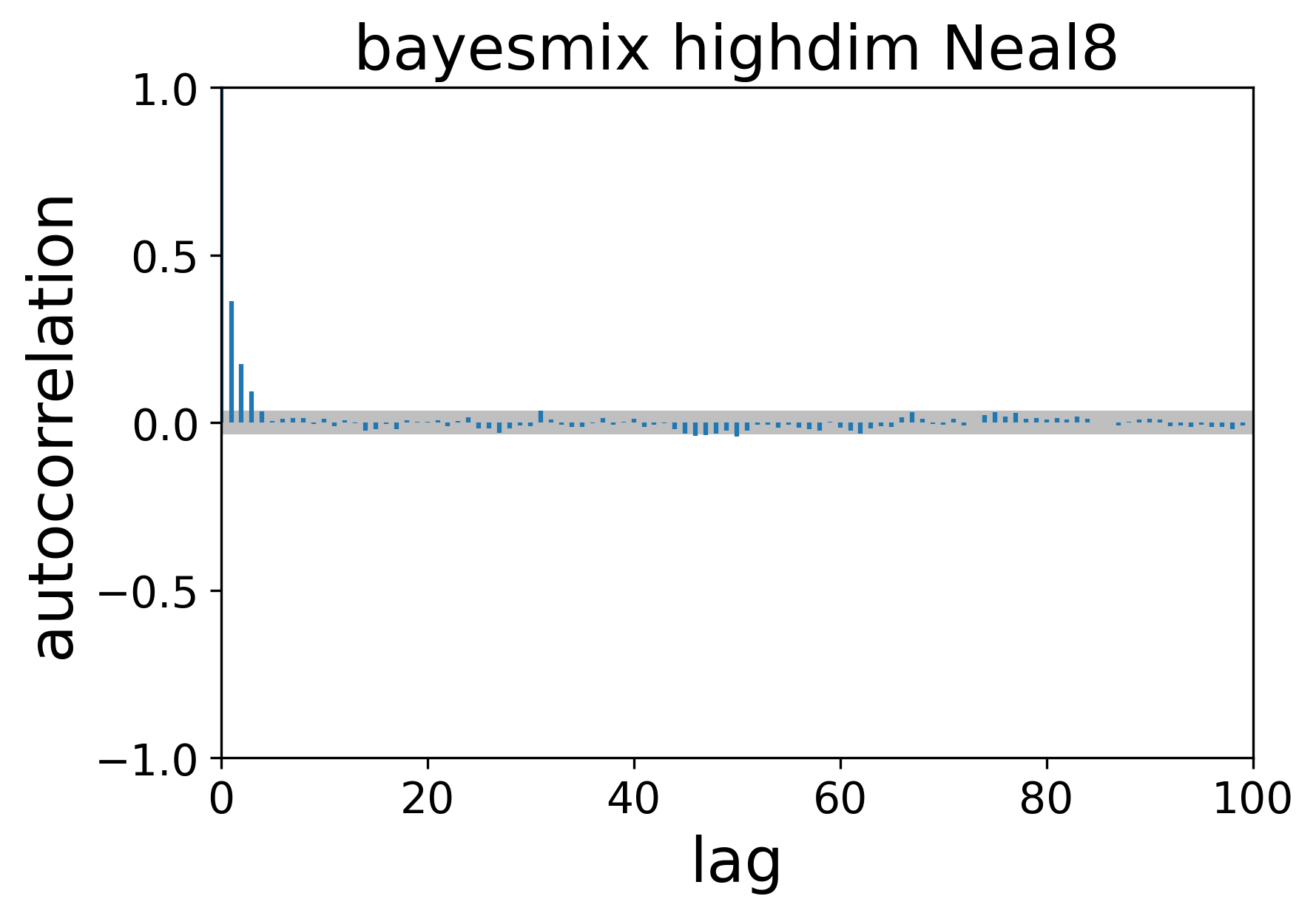}

        \includegraphics[width=0.32\textwidth]{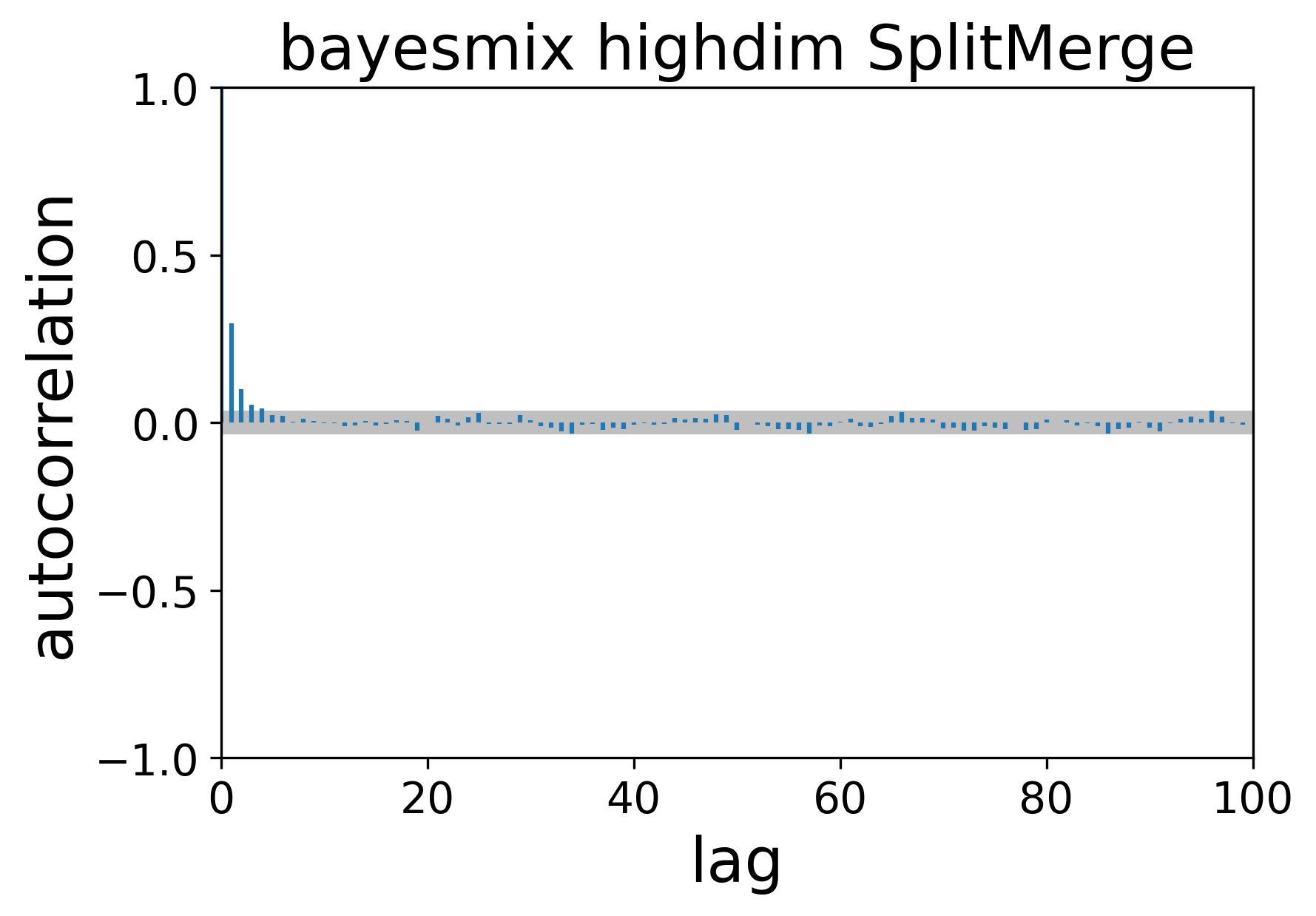}
        \includegraphics[width=0.32\textwidth]{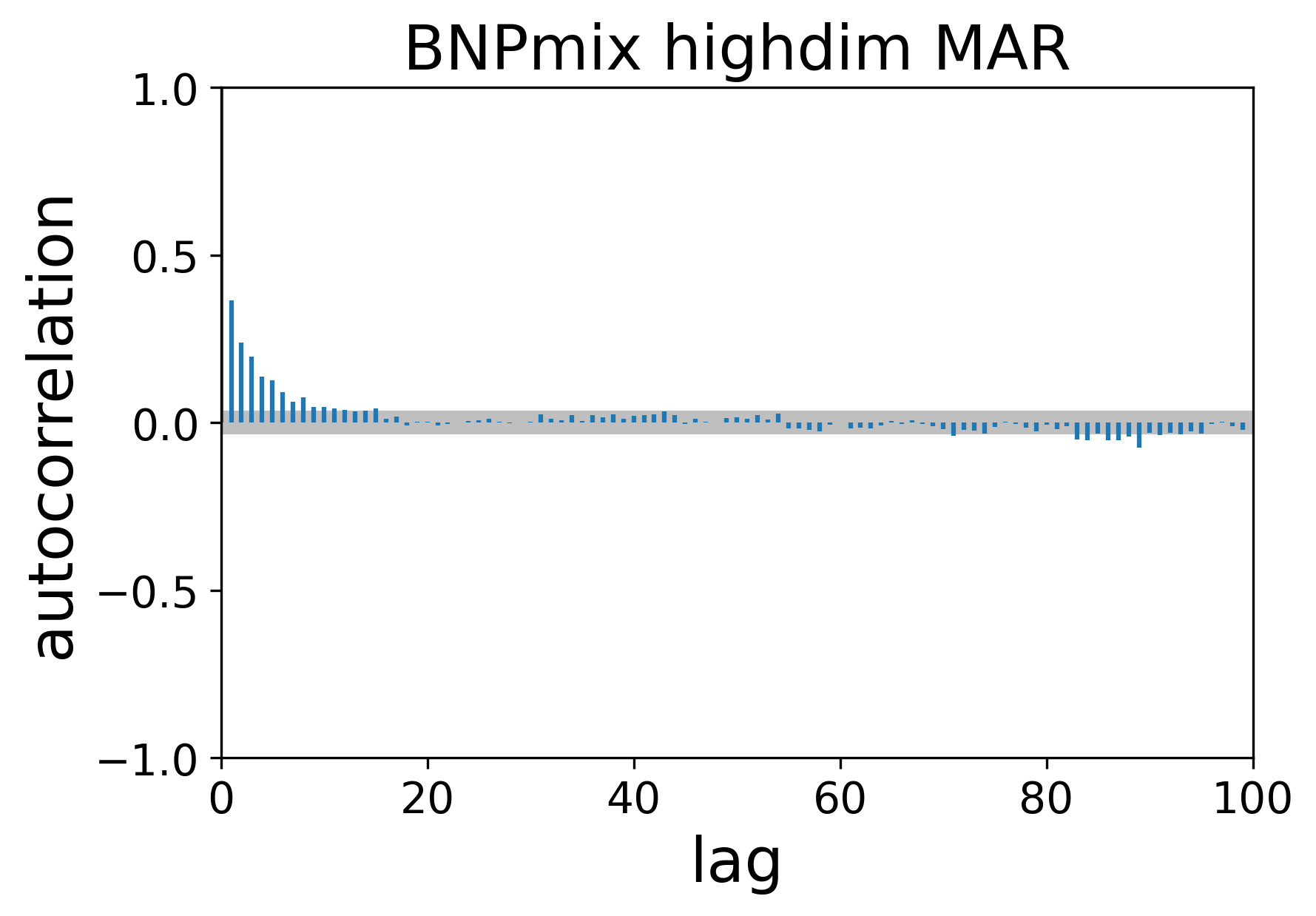}
        \includegraphics[width=0.32\textwidth]{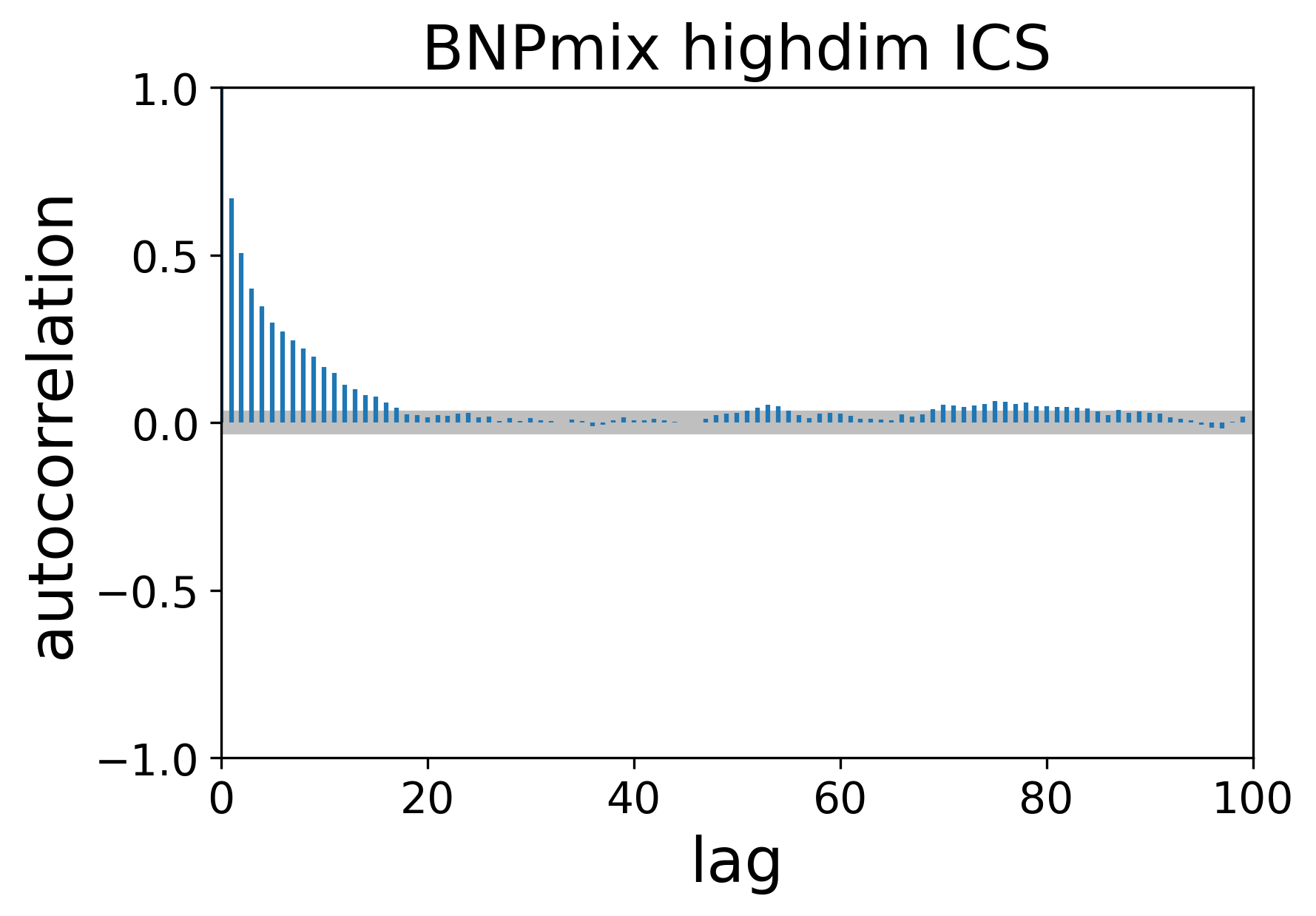}

        \caption{Comparison between autocorrelation plots on the number of clusters of the \code{galaxy} (top two rows), \code{faithful} (middle two rows), and \code{highdim} (bottom two rows) datasets}
        \label{fig:autocorr}
    \end{figure}

Autocorrelation plots for the number of clusters for all runs are displayed in Figure~\ref{fig:autocorr}.
\bayesmix{} algorithms show better mixing properties of the MCMC chain, particularly in the bivariate \code{galaxy} case, where \bnpmix{} struggles to reduce to zero the autocorrelation for large lags.

As far as computational efficiency is concerned, we
report Effective Sample Size (ESS), running times, and ESS-over-time ratio of the MCMC simulations for the above tests in Tables~\ref{tab:metrics-galaxy}, \ref{tab:metrics-faithful}, and \ref{tab:metrics-highdim}.
ESS measures the quality of a chain in terms of equivalent, hypothetical sample size of independent observations. %; therefore the higher the number, the better.
All \bayesmix{} algorithms perform much better than \bnpmix{} ones in terms of ESS while achieving comparable or lower running times.
\code{Neal2}, i.e. the same algorithm as \bnpmix{}'s \code{mar}, and \code{Neal3} stand out as being particularly efficient as quantified by the three  metrics, especially as the datapoint dimension grows larger (\code{faithful} and \code{highdim}).

\begin{figure}
\begin{table}[H]
    \centering
    \setlength{\tabcolsep}{1em}
    \renewcommand{\arraystretch}{1.5}
    \begin{tabular}{|c|c|c|c|c|}
        \hline
        & algorithm & ESS & time & ESS/time \\
        \hline \multirow{2}{*}{\bnpmix{}}
        & \code{mar} &         338.562 & 0.827 & 409.469 \\
        & \code{ics} &         162.128 & 0.842 & 192.438 \\
        \hline \multirow{4}{*}{\bayesmix{}}
        & \code{Neal2} &       337.467 & 0.370 & 912.073 \\
        & \code{Neal3} &       340.332 & 0.611 & 557.009 \\
        & \code{Neal8} &       191.580 & 0.589 & 325.263 \\
        & \code{SplitMerge} &  400.551 & 1.218 & 328.860 \\
        \hline
    \end{tabular}
    \caption{Comparison of metrics for the \code{galaxy} dataset}
    \label{tab:metrics-galaxy}
\end{table}

\begin{table}[H]
    \centering
    \setlength{\tabcolsep}{1em}
    \renewcommand{\arraystretch}{1.5}
    \begin{tabular}{|c|c|c|c|c|}
        \hline
        & algorithm & ESS & time & ESS/time \\
        \hline \multirow{2}{*}{\bnpmix{}}
        & \code{mar} &         36.288 &  3.733 &  9.721 \\
        & \code{ics} &         15.499 &  1.949 &  7.954 \\
        \hline \multirow{4}{*}{\bayesmix{}}
        & \code{Neal2} &       80.648 &  1.823 & 44.239 \\
        & \code{Neal3} &       394.709 &  4.796 & 82.300 \\
        & \code{Neal8} &       139.419 &  5.746 & 24.264 \\
        & \code{SplitMerge} &  217.788 & 12.278 & 17.738 \\
        \hline
    \end{tabular}
    \caption{Comparison of metrics for the \code{faithful} dataset}
    \label{tab:metrics-faithful}
\end{table}

\begin{table}[H]
    \centering
    \setlength{\tabcolsep}{1em}
    \renewcommand{\arraystretch}{1.5}
    \begin{tabular}{|c|c|c|c|c|}
        \hline
        & algorithm & ESS & time & ESS/time \\
        \hline \multirow{2}{*}{\bnpmix{}}
        & \code{mar} &         978.471 & 1063.740 &  0.920 \\
        & \code{ics} &         426.749 &   47.084 &  9.064 \\
        \hline \multirow{4}{*}{\bayesmix{}}
        & \code{Neal2} &       1578.956 &   44.866 & 35.193 \\
        & \code{Neal3} &       1861.819 &  166.151 & 11.206 \\
        & \code{Neal8} &       1617.569 &  296.635 &  5.453 \\
        & \code{SplitMerge} &  1865.773 &  870.494 &  2.143 \\
        \hline
    \end{tabular}
    \caption{Comparison of metrics for the \code{highdim} dataset}
    \label{tab:metrics-highdim}
\end{table}
\end{figure}

As a final example for this comparison, we have simulated ten-dimensional datapoints from a Gaussian mixture with two well- separated components (with equal weights).  
As for \code{highdim}, the sample size is 10,000.
All algorithms in \bayesmix{} but \code{Neal2} have been able to correctly distinguish the two clusters, whereas \bnpmix{} failed to do so, identifying only one.
The four- and ten-dimensional examples show that \bayesmix{} has a scalable approach that works even with large, high-dimensional datasets.

\section{Topics for expert users}
\label{sec:nerd_stuff}

The goal of this section is to give an example on how new users can extend the library by implementing a new \code{Mixing} or \code{Hierarchy}.
To do so, the \proglang{C++} code structure and the APIs of each base class must be explained  in greater detail.

We  give more details on the main building blocks in \bayesmix{}.
We follow an object-oriented approach and we adopt a combination of runtime and compile-time polymorphism based on inheritance and templates, using the so called curiously recurring template pattern (CRTP), as explained in Sections~\ref{sec:mixing} and \ref{sec:hier}.

\subsection[The Mixing module]{The \code{Mixing} module}\label{sec:mixing}

As  previously mentioned, a \code{Mixing} represents the prior distribution over the weights $\bm w$ and the  associated EPPF.
The \code{AbstractMixing} class defines the following API:
\begin{minted}{c++}
class AbstractMixing {
 public:
  virtual void initialize() = 0;
  
  virtual double get_mass_existing_cluster(
      const unsigned int n, const bool log, const bool propto,
      std::shared_ptr<AbstractHierarchy> hier,
      const Eigen::RowVectorXd &covariate=Eigen::RowVectorXd(0));

  virtual double get_mass_new_cluster(
      const unsigned int n, const bool log, const bool propto,
      const unsigned int n_clust,
      const Eigen::RowVectorXd &covariate=Eigen::RowVectorXd(0));
      
  virtual Eigen::VectorXd get_mixing_weights(
      const bool log, const bool propto,
      const Eigen::RowVectorXd &covariate = Eigen::RowVectorXd(0));

  virtual void update_state(
      const std::vector<std::shared_ptr<AbstractHierarchy>> &unique_values,
      const std::vector<unsigned int> &allocations) = 0;
};
\end{minted}
In addition to these methods, \code{AbstractMixing} defines input-output functionalities discussed in Section~\ref{sec:io}.

The \code{get_mass_existing_cluster()} and \code{get_mass_new_cluster()} methods evaluate the EPPF $\Phi$.
Specifically, \code{get_mass_existing_cluster()} evaluates $\Phi(n_1, \ldots, n_h+1, \ldots, n_k) = f_1(n_h + 1, n, \theta)$ for a given $h$, while \code{get_mass_new_cluster()} evaluates $\Phi(n_1, \ldots, n_h, \ldots, n_k + 1) = f_2(k, n, \theta)$ as defined in \eqref{eq:eppf_gibbs}.
Instead, \code{get_mixing_weights()} returns the vector of weights $\bm w$.
Both methods used to evaluate the EPPF take as input the number \code{n} of observations in the model, as well as two boolean flags (\code{propto}, \code{log}) specifying if the result must be returned up to a proportionality constant and in log-scale. The \code{get_mass_existing_cluster()} method also receives a pointer to the \code{Hierarchy} the cluster represents.
Note that the three methods take as input a vector of covariates, which is the empty vector {by default} and can be used to define \emph{dependent} mixture models, 
for instance, by assuming the dependency logit stick breaking prior implemented in \code{LogitSBMixing}. 

The \code{update_state()} method allows the child classes to assume hyperpriors on all the parameters. The \code{update_state()} method is used to sample parameters $\bm w, m$ and additional hyperparameters from their full conditional.

%\vspace{0.3cm}
\smallskip
Child classes do not inherit directly from \code{AbstractMixing}, but rather from a template class which in turn inherits from \code{AbstractMixing}, in the following way:
\begin{minted}{c++}
template <class Derived, typename State, typename Prior>
class BaseMixing : public AbstractMixing {
...
}
\end{minted}
The \code{BaseMixing} class allows for more flexible code since it is templated over two objects representing the \code{State} and the \code{Prior}. For instance, in the case of a Pitman-Yor process, the state is defined as:
\begin{minted}{c++}
namespace PitYor {
  struct State {
    double strength, discount;
  };
};
\end{minted}
but more complex objects can be used as well.
Moreover, \code{BaseMixing} implements several virtual methods from the \code{AbstractMixing} class, so that end users only need to focus on the code that is specific to a given model.
For instance, a \emph{marginal} mixing such as \code{DirichletProcess} only needs to implement the following methods:
\begin{minted}{c++}
void update_state(
      const std::vector<std::shared_ptr<AbstractHierarchy>> &unique_values,
      const std::vector<unsigned int> &allocations) override;

double mass_existing_cluster(
    const unsigned int n, const bool log, const bool propto,
    std::shared_ptr<AbstractHierarchy> hier) const override;

double mass_new_cluster(
    const unsigned int n, const bool log, const bool propto, 
    const unsigned int n_clust) const override;
\end{minted}
and some input-output functionalities. Instead, a \emph{conditional} mixing such as \code{TruncatedSBMixing} implements the following functions:
\begin{minted}{c++}
void update_state(
      const std::vector<std::shared_ptr<AbstractHierarchy>> &unique_values,
      const std::vector<unsigned int> &allocations) override;

Eigen::VectorXd get_weights(const bool log, const bool propto) const override;
\end{minted}

\subsection[The Hierarchy module]{The \code{Hierarchy} module}\label{sec:hier}

 The \code{Hierarchy} module represents the Bayesian model
\begin{equation}
\label{eq:hier_model}
    \begin{aligned}
            y_j \mid \tau &\iid f(\cdot \mid \tau), \quad j=1,\ldots,l \\
            \tau & \sim G_0
    \end{aligned}
\end{equation}
Where $f(\cdot \mid \cdot)$ is the mixture component and $G_0$ the base measure.
Given the model \eqref{eq:hier_model}, we are interested in: ($i$) evaluating the (log) likelihood function $f(x \mid \tau)$ for a given $x$, ($ii$) sampling from the prior model $\tau \sim G_0$, and ($iii$) sampling from the full conditional of $\tau \mid y_1, \ldots, y_\ell$.
Each of these goals is delegated to a different class, namely the \code{Likelihood}, the \code{PriorModel}, and the \code{Updater}.
Then a \code{Hierarchy} class is in charge of making \code{Likelihood}, \code{PriorModel}, and \code{Updater} communicate with each other and provides a common API for all possible models.

The choice of separating \code{Likelihood}, \code{PriorModel}, and \code{Updater} allows for great flexibility.
In fact, we could have different \code{Hierarchy} classes that employ the same \code{Likelihood} but a different \code{PriorModel}.
Moreover, different \code{Updater}s can be used. If the model is conjugate or semi-conjugate,  a specific \code{SemiConjugateUpdater} is   usually preferred.  If this is not the case, we provide off-the-shelf  \code{RandomWalkUpdater} and \code{MALAUpdater} that implement a random-walk Metropolis-Hastings move or a Metropolis-adjusted Langevin algorithm move, which can be used for any combination of \code{Likelihood} and \code{PriorModel}.
 As a consequence, users do not need to code an \code{Updater} if they want to implement a new model.

Throughout this section, we consider the illustrative example where $\tau = (\mu, \sigma^2)$, $f(\cdot \mid \tau) = \mathcal{N}(\cdot \mid \mu, \sigma^2)$ is the univariate Gaussian density and $G_0(\mu, \sigma^2) = \mathcal{N}(\mu \mid \mu_0, \sigma^2/\lambda) IG(\sigma^2 \mid a, b)$ is the Normal-inverse-Gamma distribution.

The \code{Hierarchy} module and all its sub-modules (\code{Likelihood}, \code{PriorModel}, \code{State} and \code{Updater}) achieve runtime polymorphism through an abstract interface (which establishes which operations can be performed by the end user) and employing the Curiously Recurring Template Pattern \citep[CRTP][]{crtp}.

Let us explain the structure in more detail, starting with the \code{Hierarchy} module. First, an \code{AbstractHierarchy} defines the following API:
\begin{minted}{c++}
class AbstractHierarchy {
 public:
  double get_like_lpdf(
      const Eigen::RowVectorXd &datum,
      const Eigen::RowVectorXd &covariate) const;
      
  virtual void sample_prior() = 0;
  
  virtual void sample_full_cond(bool update_params) = 0;
  
  virtual void add_datum(
      const int id, const Eigen::VectorXd &datum,
      const bool update_params, const Eigen::VectorXd &covariate) = 0;
      
  virtual void remove_datum(
      const int id, const Eigen::VectorXd &datum,
      const bool update_params, const Eigen::VectorXd &covariate)) = 0;
};
\end{minted}

 In the code above, \code{get_like_lpdf()} evaluates the likelihood function $f(y \mid \tau)$ for a given datapoint, \code{sample_prior()} samples from $G_0$, and \code{add_datum()} (\code{remove_datum()}) are called when allocating (removing) a datum from the current cluster.

As in the case of \code{Mixing}s, child classes inherit from a template class with respect to the \code{Likelihood} and the \code{PriorModel} from the \code{BaseHierarchy} class. 
Most of the methods in the API are implemented in this class. 
Thus, coding a new hierarchy is extremely simple within this framework, since only very few methods need to be implemented from scratch. 
All the hierarchies available so far inherit from this class and are reported in Table~\ref{tab:hiers}.

\subsubsection[The Likelihood sub-module]{The \code{Likelihood} sub-module}\label{sec:like}
The \code{Likelihood} sub-module represents the likelihood  we have assumed for the data in a given cluster. Each \code{Likelihood} class represents the sampling model
\begin{equation*}
    y_1, \ldots, y_k \mid \bm{\tau} \iid f(\cdot \mid \bm{\tau})
\end{equation*}
for a specific choice of the probability density function $f$.

In principle, the \code{Likelihood} classes are responsible only of evaluating the log-likelihood function given a specific choice of parameters $\bm \tau$. 
Therefore, a simple inheritance structure would seem appropriate.
However, the nature of the parameters $\bm \tau$ can be very different across different models (think for instance of the difference between the univariate normal and the multivariate normal paramters).
As such, we again employ CRTP to manage the polymorphic nature of \code{Likelihood} classes. 

The \code{AbstractLikelihood} class provides the following common API: 
\begin{minted}{c++}
class AbstractLikelihood {
 public:
  double lpdf(
      const Eigen::RowVectorXd &datum,
      const Eigen::RowVectorXd &covariate = Eigen::RowVectorXd(0)) const;
      
  virtual Eigen::VectorXd lpdf_grid(
      const Eigen::MatrixXd &data,
      const Eigen::MatrixXd &covariates = Eigen::MatrixXd(0, 0)) const = 0;

  virtual double cluster_lpdf_from_unconstrained(
      Eigen::VectorXd unconstrained_params) const;

  virtual stan::math::var cluster_lpdf_from_unconstrained(
      Eigen::Matrix<stan::math::var, Eigen::Dynamic, 1> unconstrained_params)
      const;

  virtual bool is_multivariate() const = 0;

  virtual bool is_dependent() const = 0;

  virtual void add_datum(
      const int id, const Eigen::RowVectorXd &datum,
      const Eigen::RowVectorXd &covariate = Eigen::RowVectorXd(0)) = 0;

  virtual void remove_datum(
      const int id, const Eigen::RowVectorXd &datum,
      const Eigen::RowVectorXd &covariate = Eigen::RowVectorXd(0)) = 0;
      

  void update_summary_statistics(const Eigen::RowVectorXd &datum,
                                 const Eigen::RowVectorXd &covariate,
                                 bool add);

  virtual void clear_summary_statistics() = 0;
};
\end{minted}
First of all, we require the implementation of the \code{lpdf()} and \code{lpdf\_grid()} methods, which simply evaluate the loglikelihood in a given point or in a grid of points (also in case of a \emph{dependent} likelihood, i.e., in which covariates are associated to each observation).
The \code{cluster\_lpdf\_from\_unconstrained()} method allows the evaluation of the likelihood of the whole cluster starting from the vector of unconstrained parameters.  
This is a key method which is only needed if a Metropolis-like updater is used.
Observe that the \code{AbstractLikelihood} class provides two such methods, one returning a \code{double} and one returning a \code{stan::math::var}. 
The latter is used to automatically compute the gradient of the likelihood via Stan's automatic differentiation, if needed.
In practice, users do not need to implement both methods separately, and can implement only one templated method; see the \code{UniNormLikelihood} example below.
The \code{add\_datum()} and \code{remove\_datum()} methods manage the insertion and deletion of a data point in the given cluster, and update the summary statistics associated with the likelihood using the \code{update\_summary\_statistics()} method.
Summary statistics (when available) are used to evaluate the likelihood function on the whole cluster, as well as to perform the posterior updates of $\bm \tau$. 
This usually gives a substantial speed-up.

Given this API, we define the \code{BaseLikelihood} class, which is a template class with respect to itself (thus enabling CRTP) and a \code{State}. The latter is a class which stores the parameters $\bm{\tau}$  and eventually manages the transformation in its unconstrained form (for Metropolis updaters), if any.
The \code{BaseLikelihood} class is declared as follows:
\begin{minted}{c++}
template <class Derived, typename State> 
    class BaseLikelihood : public AbstractLikelihood
\end{minted}
This class implements methods that are common to all the likelihoods, in order to minimize the code that end users need to implement. 
Note that every concrete implementation of a likelihood model inherits from such a class.
The following likelihoods are {currently} implemented in \bayesmix{}:
\begin{enumerate}
    \item \code{UniNormLikelihood}, that is $y \mid \mu, \sigma^2 \sim \mathcal{N}(\mu,\sigma^2)$, $\mu \in \mathbb{R}$, $\sigma^2 > 0$.
    \item \code{MultiNormLikelihood}, that is $y \mid \mu, \Sigma \sim \mathcal{N}_d(\mu, \Sigma)$, $\mu \in \mathbb{R}^d$, $\Sigma$ a symmetric and positive definite covariance matrix.
    \item \code{FALikelihood}, that is $y \mid \mu, \Sigma \sim \mathcal{N}_d(\mu, \Sigma + \Lambda \Lambda^\top)$, $\mu \in \mathbb{R}^d$, $\Sigma = \text{diag}(\sigma^2_1, \ldots, \sigma^2_d)$, $\sigma^2_j > 0$, $\Lambda$ a $d \times p$ matrix (usually $p \ll d$, hence the name factor-analyzer likelihood).
    \item \code{LinRegUniLikelihood}, that is $y \mid \beta, \sigma^2 \sim \mathcal{N}(x^\top \beta, \sigma^2)$, $\beta \in \mathbb{R}^d$, $\sigma > 0$. Here $x$ is a vector of covariates, meaning that this hierarchy is \emph{dependent}.
    \item \code{UniLapLikelihood}, that is $y \mid \mu, \lambda \sim \mathrm{Laplace}(\mu, \lambda)$, $\mu \in \mathbb{R}$, $\lambda > 0$.
\end{enumerate}

We report the code for \code{UniNormLikelihood} as an illustrative example:
\begin{minted}{c++}
class UniNormLikelihood
     : public BaseLikelihood<UniNormLikelihood, State::UniLS> {
  public:
   UniNormLikelihood() = default;
   
   ~UniNormLikelihood() = default;
   
   bool is_multivariate() const override { return false; };
   
   bool is_dependent() const override { return false; };
   
   void clear_summary_statistics() override;

   template <typename T>
   T cluster_lpdf_from_unconstrained(
       const Eigen::Matrix<T, Eigen::Dynamic, 1> &unconstrained_params) const;

  protected:
   double compute_lpdf(const Eigen::RowVectorXd &datum) const override;
   
   void update_sum_stats(const Eigen::RowVectorXd &datum, bool add) override;

   double data_sum = 0;
   
   double data_sum_squares = 0;
 };
\end{minted}

\subsubsection[The PriorModel sub-module]{The \code{PriorModel} sub-module}
\label{sec:prior}
This sub-module represents the prior for the parameters in the likelihood,  i.e.
%%model for the data in the cluster, i.e.
\begin{equation*}
    \bm{\tau} \sim G_{0}
\end{equation*}
with $G_{0}$ being a suitable prior on the parameters space. We also allow for more flexible priors adding further level of randomness (i.e. the hyperprior) on the parameter characterizing $G_{0}$. Similarly to the case of \code{Likelihood} sub-module, we need to rely on a design pattern that can manage a wide variety of specifications. We rely {once more} on the CRTP approach, thus defining an API via a pure virtual class: \code{AbstractPriorModel}, which collects the methods each class should implement. This class is defined as follows:
\begin{minted}{c++}
class AbstractPriorModel {
 public:
  
  virtual double lpdf(const google::protobuf::Message &state_) = 0;
  
  virtual double lpdf_from_unconstrained(
      Eigen::VectorXd unconstrained_params) const;

  virtual stan::math::var lpdf_from_unconstrained(
      Eigen::Matrix<stan::math::var,Eigen::Dynamic,1> unconstrained_params) const;

  virtual std::shared_ptr<google::protobuf::Message> sample(
      ProtoHypersPtr hier_hypers = nullptr) = 0;

  virtual void update_hypers(
      const std::vector<bayesmix::AlgorithmState::ClusterState> &states) = 0;
};
\end{minted}
The \code{lpdf()} and \code{lpdf\_from\_unconstrained()} methods evaluate the log-prior density function at the current state $\bm \tau$ or its unconstrained representation.
In particular, \code{lpdf\_from\_unconstrained()} is needed by Metropolis-like updaters; see below for further details.
The \code{sample()} method generates a draw from the prior distribution. If \code{hier_hypers} is \code{nullptr}, the prior hyperparameter values are used.
To allow sampling from the full conditional distribution in case of semi-congugate hierarchies, we introduce the \code{hier\_hypers} parameter, which is a pointer to a \code{Protobuf} message storing the hierarchy hyperaprameters to use for the sampling.
The \code{update\_hypers()} method updates the prior hyperparameters, given the vector of all cluster states.

Given the API, we define the \code{BasePriorModel} class, which is declared as:
\begin{minted}{c++}
template <class Derived, class State, typename HyperParams, typename Prior>
class BasePriorModel : public AbstractPriorModel
\end{minted}
Such a class is derived from \code{AbstractPriorModel}. It is a template class with respect to itself (for CRTP), a \code{State} class (which represents the parameters over which the prior is assumed) an \code{HyperParams} type (which is a simple struct that codes the parameters characterizing $G_{0}$) and a \code{Prior} (which codes hierarchical priors for the $G_0$ parameters for more flexible and robust prior models).
Like in previous sub-modules, this class manages code exceptions and implements general methods.
Every concrete implementation of a prior model must be defined as an inherited class of \code{BasePriorModel}.
The library currently supports the following priors:
\begin{enumerate}
    \item \code{NIGPriorModel} $\mu \mid \sigma^2 \sim \mathcal{N}(\mu_0, \sigma^2 / \lambda)$, $\sigma^2 \sim IG(a, b)$.
    \item \code{NxIGPriorModel} $\mu \sim \mathcal{N}(\mu_0, \sigma_0^2)$, $\sigma^2 \sim IG(a, b)$.
    \item \code{NWPriorModel} $\mu \mid \Sigma \sim \mathcal{N}(\mu_0, \Sigma/\lambda)$, $\Sigma \sim IW(\nu_0, \Psi_0)$.
    \item \code{MNIGPriorModel} $\beta \mid \sigma^2 \sim N_{p}(\mu, \sigma^2 \Lambda^{-1})$, $\sigma^2 \sim IG(a, b)$
    \item \code{FAPriorModel} $\mu \sim \mathcal{N}_p(\widetilde \mu, \psi I)$, $\Lambda \sim \mbox{DL}(\alpha)$, $\Sigma = \mbox{diag}(\sigma_1, \ldots, \sigma_p)$, $\sigma_j \iid IG(a, b)$, $j=1, \ldots, p$, where $\mbox{DL}$ is the Dirichlet-Laplace distribution in \cite{bhattacharya2015dirichlet}.
\end{enumerate} 
As an example, we report the implementation of the \code{NIGPriorModel} here below:
\begin{minted}{c++}
class NIGPriorModel : public BasePriorModel<
    NIGPriorModel, State::UniLS, Hyperparams::NIG, bayesmix::NNIGPrior> {
 public:
  using AbstractPriorModel::ProtoHypers;

  using AbstractPriorModel::ProtoHypersPtr;

  NIGPriorModel() = default;

  ~NIGPriorModel() = default;

  double lpdf(const google::protobuf::Message &state_) override;

  template <typename T>
  T lpdf_from_unconstrained(
      const Eigen::Matrix<T, Eigen::Dynamic, 1> &unconstrained_params) const {
    Eigen::Matrix<T, Eigen::Dynamic, 1> constrained_params =
        State::uni_ls_to_constrained(unconstrained_params);
    T log_det_jac = State::uni_ls_log_det_jac(constrained_params);
    T mean = constrained_params(0);
    T var = constrained_params(1);
    T lpdf = stan::math::normal_lpdf(mean, hypers->mean,
                                     sqrt(var / hypers->var_scaling)) +
             stan::math::inv_gamma_lpdf(var, hypers->shape, hypers->scale);

    return lpdf + log_det_jac;
  };
  
  State::UniLS sample(ProtoHypersPtr hier_hypers = nullptr) override;
  
  void update_hypers(const std::vector<bayesmix::AlgorithmState::ClusterState>
                         &states) override;
  
  void set_hypers_from_proto(
      const google::protobuf::Message &hypers_) override;
  
  std::shared_ptr<bayesmix::AlgorithmState::HierarchyHypers> get_hypers_proto()
      const override;

 protected:
  void initialize_hypers() override;
};
\end{minted}

\subsubsection[The Updater sub-module]{The \code{Updater} sub-module}\label{sec:updater}
The \code{Updater} module implements the machinery to provide a sampling from the full conditional distribution of a given hierarchy. Again, we rely on CRTP and define the API in the \code{AbstractUpdater} class as follows:
\begin{minted}{c++}
class AbstractUpdater {
 public:
  virtual bool is_conjugate() const;
  
  virtual void draw(AbstractLikelihood &like, AbstractPriorModel &prior,
                    bool update_params) = 0;s
};
\end{minted}
Here \code{is\_conjugate()} declares whether the updater is meant to be used for a semi-conjugate hierarchy.
The \code{draw} method is the key method of every updater: it receives \code{like} and \code{prior} as input, and updates the \code{State} (which is stored inside the \code{Likelihood}) by sampling it from conditional distribution $\bm \tau \mid y_1, \ldots, y_h$, where the $y_j$'s are the data associated to one specific cluster. 
As already mentioned, when \eqref{eq:hier_model} is semi-conjugate, problem-specific updaters can be easily implemented by inheriting from the \code{SemiConjugateUpdater}; see, for instance, the code below.
\begin{minted}{c++}
class NNIGUpdater: public SemiConjugateUpdater<UniNormLikelihood, NIGPriorModel> {
  public:
   NNIGUpdater() = default;
   ~NNIGUpdater() = default;

   bool is_conjugate() const override { return true; };

   ProtoHypers compute_posterior_hypers(AbstractLikelihood &like,
                                        AbstractPriorModel &prior) override;
};
\end{minted}
In particular, note that this class does not implement any \code{draw()} method. In fact, since the model is semi-conjugate, we  exploit the \code{PriorModel} draw function but using updated parameters, which are computed by the \code{compute_posterior_hypers()} method.

If the model is not semi-conjugate, we suggest using \code{RandomWalkUpdater} or \code{MALAUpdater}, which sample from the full conditional distribution of $\bm \tau$ using a Metropolis-Hastings move.
In this case, the following methods must be implemented in the \code{Likelihood} class:
\begin{minted}{c++}
template <typename T>
T cluster_lpdf_from_unconstrained(
       const Eigen::Matrix<T, Eigen::Dynamic, 1> &unconstrained_params) const;
\end{minted}
while the prior should implement the following:
\begin{minted}{c++}
template <typename T>
T lpdf_from_unconstrained(
       const Eigen::Matrix<T, Eigen::Dynamic, 1> &unconstrained_params) const;
\end{minted}
For instance,  when $f$ is the univariate Gaussian density, the unconstrained parameters are $(\mu, \log(\sigma^2))$. 
To evaluate the likelihood, it is sufficient to transform $\log(\sigma^2)$ using the exponential function. Instead, to evaluate the prior, one should take care of the correction in the density function due to the change of variables.

\subsubsection[The State sub-module]{The \code{State} sub-module}\label{sec:state}

\code{States} are classes used to store  parameters $\tau_h$'s of every mixture component.
Their main purpose is to handle serialization and de-serialization of the state; see also Section \ref{sec:io}.
Moreover, they allow to go from the \code{constrained} to the \code{unconstrained} representation of the parameters (and viceversa) and compute the associated determinant of the Jacobian appearing in the change of density formula. All states inherit from a \code{BaseState}:
\begin{minted}{c++}
class BaseState {
 public:
  int card;
  using ProtoState = bayesmix::AlgorithmState::ClusterState;

  virtual Eigen::VectorXd get_unconstrained() { throw std::runtime_error("..."); }
  virtual void set_from_unconstrained(const Eigen::VectorXd &in) { 
    throw std::runtime_error("..."); }
  virtual double log_det_jac() { throw std::runtime_error("..."); }

  virtual void set_from_proto(const ProtoState &state_, bool update_card) = 0;
  virtual ProtoState get_as_proto() const = 0;
  std::shared_ptr<ProtoState> to_proto() const {
    return std::make_shared<ProtoState>(get_as_proto());
  }
};
\end{minted}
Depending on the chosen \code{Updater}, the methods \code{get_unconstrained()}, \code{set_from_unconstrained()} and \code{log_det_jac()} might never be called. 
Therefore, we do not force users to implement them.
Instead, the \code{set_from_proto()} and \code{get_as_proto()} are fundamental as they allow the interaction with Google's Protocol Buffers library; see Section \ref{sec:io} for more detail.

\subsection[The Algorithm module]{The \code{Algorithm} module}

\code{Mixing} and \code{Hierarchy} classes are combined together by an \code{Algorithm}.
Algorithms are direct implementation of MCMC samplers, such as Neal's Algorithm 2/3/8  and the blocked Gibbs sampler from \cite{ishwaran2001gibbs}.
All algorithms must inherit from the \code{BaseAlgorithm} class:
\begin{minted}{c++}
class BaseAlgorithm {
 protected:
  Eigen::MatrixXd data;
  Eigen::MatrixXd hier_covariates;
  Eigen::MatrixXd mix_covariates;
  
  std::vector<unsigned int> allocations;
  std::vector<std::shared_ptr<AbstractHierarchy>> unique_values;
  std::shared_ptr<BaseMixing> mixing;
 
  virtual void sample_allocations() = 0;
  virtual void sample_unique_values() = 0;

  virtual void step() {}

 public:
  void run(BaseCollector *collector);
  virtual Eigen::MatrixXd eval_lpdf(
      BaseCollector *const collector, const Eigen::MatrixXd &grid,
      const Eigen::MatrixXd &hier_covariates = Eigen::MatrixXd(0, 0),
      const Eigen::MatrixXd &mix_covariates = Eigen::MatrixXd(0, 0)) = 0;
};
\end{minted}
The \code{Algorithm}  class saves the data and (optionally) two set of covariates: \code{hier_covariates} and \code{mix_covariates}. Therefore, it is trivial to extend the code to more general models 
to accommodate for covariate-dependent likelihoods and/or mixings.
Moreover, the \code{Algorithm} also stores the cluster allocation variables (\code{allocations}), the hierachies representing the mixture components (\code{unique_values}) and the mixing (\code{mixing}).
The last two objects are stored through pointers to the corresponding base class, to achieve runtime polymorphism. 

The basic method from \code{Algorithm} is \code{step()} which performs a Gibbs sampling step calling the appropriate update methods for all the blocks of the model.
A \code{run()} method is used to run the MCMC chain, i.e. \code{run()} calls \code{step()} for a user-specified number of iterations, possibly discarding an initial burn-in phase.
The goal of MCMC simulations is to \emph{collect} samples from the posterior distribution, which must be stored  for later use. 
Hence, the \code{run()} receives as input an instance of \code{BaseCollector} which is indeed in charge of storing the visited states either in memory (RAM) or by saving in a file; see Section~\ref{sec:io} for further details.

Since one of the main goals of mixture analysis is density estimation, an \code{Algorithm} must be also able to evaluate the mixture density on a fixed grid, given the visited samples. This is achieved by the \code{eval_lpdf()} method.

 All the algorithms implemented in \bayesmix{} are listed in Table~\ref{tab:algos}.

\subsection{I/O and cross-language functionalities}\label{sec:io}

There is a final building block of \bayesmix{}, that is the management of input / output (I/O).
Most of \proglang{C++} based packages for Bayesian inference, such as \pkg{Stan} \citep{stan2019stan} and \pkg{JAGS} \citep{plummer2017jags}, rely on tabular formats to save the chains. 
Specifically, the output of an MCMC  algorithm is collected in an array where each parameter is saved in a different column  and the resulting object is then serialized in a text format (such as csv).
This approach is simple but rather restrictive, since it requires a fixed number of parameters, which is usually not our case.
Moreover, in case of non-scalar parameters (such as covariance matrices), these parameters need to be first \emph{flattened}   to be stored in a matrix and then they need to be re-built from this flattened version to  compute posterior inference.

Instead, we rely on the powerful serialization library Protocol Buffers (\url{https://developers.google.com/protocol-buffers/}) to handle I/O operations.
Specifically, this requires defining so-called \emph{messages} in a \code{.proto} file. Semantically, the declaration of a message is  alike the declaration of a \proglang{C++} struct.
For instance the following code:
\begin{minted}{c++}
message UniLSState {
  double mean = 1;
  double var = 2;
}
\end{minted}
defines a message named \code{UniLSState} whose fields are two \code{double}s, \code{mean} and \code{var}. In more complex settings, other \code{Protobuf} messages can act as types for these variables.
The \code{protoc} compiler operates on these messages and transpiles them into files implementing associated classes (one per message) in a given programming language (for us, it is of course \proglang{C++}). Then, the runtime library \code{google/protobuf} can be used to serialize and deserialize these messages very efficiently.
All messages are declared in files placed in the \code{proto} folder. The transpilation into the corresponding \code{C++} classes occurs automatically when installing the \bayesmix{} library.

The state of the Markov chain can be stored in the following message:
\begin{minted}{c++}
message AlgorithmState {
  repeated ClusterState cluster_states = 1;
  repeated int32 cluster_allocs = 2 [packed = true];
  MixingState mixing_state = 3;
  int32 iteration_num = 4;
  HierarchyHypers hierarchy_hypers = 5;
}
\end{minted}
where \code{ClusterState}, \code{MixingState} and \code{HierarchyHypers} are other messages defined in the \code{proto} folder.

In our code, there are classes that are exclusively dedicated to storing the samples from the MCMC, either in memory or on file. These are called \code{Collectors} and inherit from \code{BaseCollector} that defines the API:
\begin{minted}{c++}
class BaseCollector {
 public:
  virtual void start_collecting() = 0;

  virtual void finish_collecting() = 0;

  bool get_next_state(google::protobuf::Message *out);

  virtual void collect(const google::protobuf::Message &state) = 0;

  virtual void reset() = 0;

  unsigned int get_size() const;
\end{minted}
A collector stores the entire MCMC chain in a data structure that resembles a linked list, that is, the collector knows the beginning of the chain and the current state. The function \code{get_next_state()} can be used to advance to the next state, while writing its values to a pointer. Instead, the algorithm calls the \code{collect()} method when a MCMC iteration must be saved.

\subsection{Extending the BayesMix library}\label{sec:expansion}

In this section, we show a concrete example of an extension of \bayesmix{}.
We consider a mixture model with $\text{Gamma}(\cdot \mid \alpha, \beta)$ kernel, where $\alpha$ is a fixed parameter, and the mixing measure over $\beta$ is a Dirichlet process  with conjugate $\text{Gamma}(\alpha_0, \beta_0)$ base measure.
We can use any of the algorithms in \bayesmix{} to sample from the posterior of this model, but we need to implement additional code in our library.

Three or four classes are needed: (i) a \code{GammaLikelihood} class representing a Gamma likelihood, (ii) a \code{GammaPriorModel} class representing a Gamma prior over the $\tau_h$'s, and (iii) a \code{GammaHierarchy} that combines \code{GammaLikelihood} and \code{GammaPriorModel}. As far as the updater is concerned, we could either use a \code{MetropolisUpdater} or implement a (iv) \code{GammaGammaUpdater} class that takes advantage of the conjugacy. In this example, we opt for the latter.

We will not cover in full detail the implementation of all the required functions, but just the core ones. The full code for this example is available at \url{https://github.com/bayesmix-dev/bayesmix/tree/master/examples}.

Since the state of each component is just $(\alpha, \beta_h)$, where $\alpha$ is fixed in our case, we can use the \code{Protobuf} message \code{bayesmix::AlgorithmState::ClusterState::general_state} to save it. That is, we save each $(\alpha, \beta_h)$ in a \code{Vector} of length two. This is done in the \code{geta_as_proto()} function implemented below.
For more complex hierarchies, we suggest users to create their own \code{Protobuf} messages and add them to the \code{bayesmix::AlgorithmState::ClusterState} field.

We report the code for the \code{State} and \code{GammaLikleihood} classes below:

\begin{minted}{C++}
namespace State { class Gamma: public BaseState {
 public:
  double shape, rate;
  using ProtoState = bayesmix::AlgorithmState::ClusterState;

  ProtoState get_as_proto() const override {
    ProtoState out;
    out.mutable_general_state()->set_size(2);
    out.mutable_general_state()->mutable_data()->Add(shape);
    out.mutable_general_state()->mutable_data()->Add(rate);
    return out;
  }

  void set_from_proto(const ProtoState &state_, bool update_card) override {
    if (update_card) { card = state_.cardinality(); }
    shape = state_.general_state().data()[0];
    rate = state_.general_state().data()[1];
  }
};}
\end{minted}

\begin{minted}{c++}
 class GammaLikelihood : public BaseLikelihood<GammaLikelihood, State::Gamma> {
 public:
  ...
  void clear_summary_statistics() override;

 protected:
  double compute_lpdf(const Eigen::RowVectorXd &datum) const override;
  void update_sum_stats(const Eigen::RowVectorXd &datum, bool add) override;

  double data_sum = 0;
  int ndata = 0;
};

void GammaLikelihood::clear_summary_statistics() {
  data_sum = 0;
  ndata = 0;
}

double GammaLikelihood::compute_lpdf(const Eigen::RowVectorXd &datum) const {
  return stan::math::gamma_lpdf(datum(0), state.shape, state.rate);
}

void GammaLikelihood::update_sum_stats(const Eigen::RowVectorXd &datum,
                                       bool add) {
  if (add) {
    data_sum += datum(0);
    ndata += 1;
  } else {
    data_sum -= datum(0);
    ndata -= 1;
  }
}
\end{minted}

Next, we report the code for the \code{GammaPriorModel} class.
As we did for the \code{GammaLikelihood}, we do not need to write any additional \code{Protobuf} messages.
Instead, we rely on the \\
\code{HierarchyHypers::general_state} field which saves the hyperparameters $\alpha_0$ and $\beta_0$ in a \code{Vector}. 

\begin{minted}{c++}
namespace Hyperparams {
 struct Gamma {
   double rate_alpha, rate_beta;
 }; 
} 

class GammaPriorModel
    : public BasePriorModel<GammaPriorModel, State::Gamma, Hyperparams::Gamma,
                            bayesmix::EmptyPrior> {
 public:
  using AbstractPriorModel::ProtoHypers;
  using AbstractPriorModel::ProtoHypersPtr;

  GammaPriorModel(double shape_ = -1, double rate_alpha_ = -1,
                  double rate_beta_ = -1);
  ~GammaPriorModel() = default;

  double lpdf(const google::protobuf::Message &state_) override;

  State::Gamma sample(ProtoHypersPtr hier_hypers = nullptr) override;

  void update_hypers(const std::vector<bayesmix::AlgorithmState::ClusterState>
                         &states) override {
    return;
  };

  void set_hypers_from_proto(
      const google::protobuf::Message &hypers_) override;

  ProtoHypersPtr get_hypers_proto() const override;
  double get_shape() const { return shape; };

 protected:
  double shape, rate_alpha, rate_beta;
  void initialize_hypers() override;
};

/* DEFINITIONS */
GammaPriorModel::GammaPriorModel(double shape_, double rate_alpha_,
                                 double rate_beta_)
    : shape(shape_), rate_alpha(rate_alpha_), rate_beta(rate_beta_) {
  create_empty_prior();
};

double GammaPriorModel::lpdf(const google::protobuf::Message &state_) {
  double rate = downcast_state(state_).general_state().data()[1];
  return stan::math::gamma_lpdf(rate, hypers->rate_alpha, hypers->rate_beta);
}

State::Gamma GammaPriorModel::sample(
    ProtoHypersPtr hier_hypers) {
  auto &rng = bayesmix::Rng::Instance().get();
  State::Gamma out;

  auto params = (hier_hypers) ? hier_hypers->general_state()
                              : get_hypers_proto()->general_state();
  double rate_alpha = params.data()[0];
  double rate_beta = params.data()[1];
  out.shape = shape;
  out.rate = stan::math::gamma_rng(rate_alpha, rate_beta, rng);
  return out;
}

void GammaPriorModel::set_hypers_from_proto(
    const google::protobuf::Message &hypers_) {
  auto &hyperscast = downcast_hypers(hypers_).general_state();
  hypers->rate_alpha = hyperscast.data()[0];
  hypers->rate_beta = hyperscast.data()[1];
};

GammaPriorModel::ProtoHypersPtr GammaPriorModel::get_hypers_proto() const {
  ProtoHypersPtr out = std::make_shared<ProtoHypers>();
  out->mutable_general_state()->mutable_data()->Add(hypers->rate_alpha);
  out->mutable_general_state()->mutable_data()->Add(hypers->rate_beta);
  return out;
};

void GammaPriorModel::initialize_hypers() {
  hypers->rate_alpha = rate_alpha;
  hypers->rate_beta = rate_beta;

  // Checks
  if (shape <= 0) {
    throw std::runtime_error("shape must be positive");
  }
  if (rate_alpha <= 0) {
    throw std::runtime_error("rate_alpha must be positive");
  }
  if (rate_beta <= 0) {
    throw std::runtime_error("rate_beta must be positive");
  }
}
\end{minted}

Finally, we implement a dedicated \code{Updater} as follows.

\begin{minted}{c++}
class GammaGammaUpdater
     : public SemiConjugateUpdater<GammaLikelihood, GammaPriorModel> {
  public:
   GammaGammaUpdater() = default;
   ~GammaGammaUpdater() = default;

   bool is_conjugate() const override { return true; };

   ProtoHypersPtr compute_posterior_hypers(
    AbstractLikelihood& like, AbstractPriorModel& prior) override {
     // Likelihood and Prior downcast
     auto& likecast = downcast_likelihood(like);
     auto& priorcast = downcast_prior(prior);

     // Getting required quantities from likelihood and prior
     int card = likecast.get_card();
     double data_sum = likecast.get_data_sum();
     double ndata = likecast.get_ndata();
     double shape = priorcast.get_shape();
     auto hypers = priorcast.get_hypers();

     // No update possible
     if (card == 0) {
       return priorcast.get_hypers_proto();
     }
     // Compute posterior hyperparameters
     double rate_alpha_new = hypers.rate_alpha + shape * ndata;
     double rate_beta_new = hypers.rate_beta + data_sum;

     // Proto conversion
     ProtoHypers out;
     out.mutable_general_state()->mutable_data()->Add(rate_alpha_new);
     out.mutable_general_state()->mutable_data()->Add(rate_beta_new);
     return std::make_shared<ProtoHypers>(out);
   }
};
\end{minted}
 
 Note that implementing this new model has required only less than 130 lines of code. In particular, the coding effort could be substantially reduced by using, for instance, the \code{RandomWalkUpdater} instead of writing a custom \code{GammaGammaUpdater}.

\section{Summary and Future Developments}
\label{sec:discussion}

In this paper, we have presented \bayesmix{}, a \proglang{C++} library for posterior inference in Bayesian (nonparametric) mixture models. 
Compared to previously available software, our library features greater flexibility and extensibility, as shown by  the modularity of our code, which makes it easy to extend our library to other mixture models.
Therefore, \bayesmix{} provides an ideal software ecosystem for computer scientists, statisticians and practitioners who need to consider complex models.
As shown by the examples, our library compares favourably to the competitor package in terms of computational efficiency and of overall quality of the output MCMC samples.

The main limitation of \bayesmix{} is also its point of strength, that is being a \proglang{C++} library.
As such, \proglang{C++} programmers can benefit from the rich language and the efficiency of the \proglang{C++} code to easily extend our library to their needs.
However, knowledge of \proglang{C++} might represent a barrier for new users.

To this end, we are currently developing the \proglang{Python} package \pkg{pybmix} (\url{https://github.com/bayesmix-dev/pybmix}), whose ultimate goal will be to allow the same degree of extensibility without knowledge of \proglang{C++}; users will be able to extend our library writing code solely in \proglang{Python}.
Of course, this causes a loss in efficiency, since \proglang{Python} is slower than \proglang{C++} and there issubstantial overhead in calling \proglang{Python} code from \proglang{C++}.
However, compared to pure \proglang{Python} implementations, we expect our approach to be faster in terms of both runtime and development time (i.e., the time required to code an MCMC algorithm).
We could certainly achieve the same goal within an \proglang{R} package, but at the moment this is not being considered.

 The latest version of our library can be found at the official Github repository at \url{https://github.com/bayesmix-dev/bayesmix}.
At the moment, our project has 14 contributors.
Any interested user or developer can easily get in touch with us through our Github repository by opening an issue or requesting new features. We welcome any contribution to \bayesmix{} and the \proglang{Python} package \pkg{pybmix}. Moreover, we would be happy to provide support to developers aiming at building an \proglang{R} package interface.

\section*{Acknowledgements}

We thank all the developers who contributed to the \bayesmix{} library and, in particular:
Matteo Bollettino, Giacomo De Carlo, Alessandro Carminati, Madhurii Gatto, Taguhi Mesropyan, Vittorio Nardi, Enrico Paglia, Giovanni Battista Pollam, Pasquale Scaramuzzino.
We are also grateful to Elena Zazzetti, who started working on this library when it was in an embryonal stage.

\bibliography{references}

\end{document}